\documentclass[review=false]{jfp-epi}

\setcitestyle{nosort} 

\usepackage[utf8]{inputenc}
\usepackage{stmaryrd}
\usepackage{empheq}
\usepackage{mathpartir}
\usepackage{polytable}
\usepackage{siunitx}
\usepackage{newunicodechar}
\usepackage{microtype}
\usepackage[many]{tcolorbox}
\usepackage{agda}
\usepackage{cleveref}
\renewcommand{\AgdaFontStyle}[1]{\scalebox{0.85}{\textsf{#1}}}

\usepackage{graphicx} 

\renewcommand{\mathsf}[1]{\scalebox{0.93}{\ensuremath{\text{\sf#1}}}}

\newcommand{\eg}{e.g.}
\newcommand{\ie}{i.e.}

\newcommand{\TOP}[1]{\ignorespaces}  
\newcommand{\DELETED}[1]{}

\definecolor{cca}{RGB}{188,26,28}
\definecolor{ccb}{RGB}{55,126,184}
\definecolor{ccc}{RGB}{255,127,0}
\definecolor{ccd}{RGB}{77,175,74}
\definecolor{cce}{RGB}{152,78,163}
\definecolor{ccf}{RGB}{247,129,191}
\definecolor{ccg}{RGB}{255,255,51}
\definecolor{cch}{RGB}{166,86,40}
\definecolor{cci}{RGB}{153,153,153}
\newcommand{\ca}{\textcolor{cca}}
\newcommand{\cb}{\textcolor{ccb}}
\newcommand{\cc}{\textcolor{ccc}}
\newcommand{\cd}{\textcolor{ccd}}
\newcommand{\ce}{\textcolor{cce}}
\newcommand{\cf}{\textcolor{ccf}}

\newcommand{\qzUC}{q_z}
\newcommand{\qsUC}{q_s}
\newcommand{\qnUC}{q_n}
\newcommand{\qz}{\cd{q_{\mathrm{z}}}}
\newcommand{\qs}{\cf{q_{\mathrm{s}}}}
\newcommand{\qn}{\ce{q_{\mathrm{n}}}}
\newcommand{\qzp}{\cd{q'_{\mathrm{z}}}}
\newcommand{\qsp}{\cf{q'_{\mathrm{s}}}}
\newcommand{\qnp}{\ce{q'_{\mathrm{n}}}}
\newcommand{\gz}{\cd{\gamma_{\mathrm{z}}}}
\newcommand{\gs}{\cf{\gamma_{\mathrm{s}}}}
\newcommand{\gn}{\ce{\gamma_{\mathrm{n}}}}

\renewcommand{\L}{\mathsf{L}}
\newcommand{\M}{\mathsf{M}}
\renewcommand{\H}{\mathsf{H}}

\newcommand{\meet}{\wedge}
\newcommand{\kStar}[1]{#1^*}
\newcommand{\tail}[1]{\mathsf{tail} \, #1}
\newcommand{\head}[1]{\mathsf{head} \, #1}

\newcommand{\nrName}{\ensuremath{\mathsf{nr}}}
\newcommand{\nr}[5]{\nrName_{#1}^{#2} \, #3 \, #4 \, #5}

\newcommand{\tProdrec}{\mathsf{Prodrec}}
\newcommand{\tUnitrec}{\mathsf{Unitrec}}
\newcommand{\tEmptyrec}{\mathsf{Emptyrec}}
\newcommand{\Prodrec}[1]{\tProdrec\,{#1}}
\newcommand{\Unitrec}[1]{\tUnitrec\,{#1}}
\newcommand{\Emptyrec}[1]{\tEmptyrec\,{#1}}

\newcommand{\neutral}[1]{\mathsf{Ne}\,#1}

\newcommand{\Type}{\mathsf{Type}}
\newcommand{\U}{\mathsf{U}}
\newcommand{\Nat}{\mathbb{N}}
\newcommand{\bN}{\mathbb{N}}
\newcommand{\PiName}[2]{\Pi^{#2}_{#1}}
\newcommand{\PiType}[4]{\Pi^{#2}_{#1} \, #3 \, #4}
\newcommand{\SigmaType}[3]{\Sigma^{#1} \, #2 \, #3}
\newcommand{\SigmaTypeX}[4][k]{\Sigma^{#2}_{#1} \, #3 \, #4}
\newcommand{\GradedSigmaTypeX}[5][k]{\Sigma^{#3}_{#1,#2} \, #4 \, #5}
\newcommand{\weakLabel}{\otimes}
\newcommand{\strongLabel}{\&}
\newcommand{\SigmaTypeW}[3]{\SigmaTypeX[\weakLabel]{#1}{#2}{#3}}
\newcommand{\SigmaTypeS}[3]{\SigmaTypeX[\strongLabel]{#1}{#2}{#3}}

\newcommand{\GradedSigmaTypeWWithoutArgs}[2]{\Sigma^{#2}_{\weakLabel,#1}}
\newcommand{\GradedSigmaTypeS}[3]{\GradedSigmaTypeX[\strongLabel]{#1}{#2}{#3}}
\newcommand{\GradedSigmaTypeSWithoutArgs}[2]{\Sigma^{#2}_{\strongLabel,#1}}
\newcommand{\Unit}{\top}
\newcommand{\UnitX}[1][k]{\Unit_{#1}}

\newcommand{\Empty}{\bot}

\newcommand{\var}[1]{\mathsf{x}_{#1}}
\newcommand{\lam}[2]{\lambda^{#1} #2}
\newcommand{\app}[3]{#1 \,^{#2} #3}

\newcommand{\zero}{\mathsf{zero}}
\newcommand{\sucName}{\mathsf{suc}}
\newcommand{\suc}[1]{\sucName{} \, #1}
\newcommand{\natrecName}{\mathsf{natrec}}
\newcommand{\natrec}[7]{\natrecName{}_{#1,#3}^{#2} \, #4 \, #5 \, #6 \, #7}
\newcommand{\numeral}[1]{\underline{#1}}

\newcommand{\pair}[2]{(#1, #2)}
\newcommand{\pairX}[3][k]{(#2, #3)_{#1}}
\newcommand{\pairW}[2]{\pairX[\weakLabel]{#1}{#2}}
\newcommand{\pairS}[2]{\pairX[\strongLabel]{#1}{#2}}
\newcommand{\pairM}[3]{(^{#3}#1, #2)}
\newcommand{\pairMX}[4][k]{(^{#4}#2, #3)_{#1}}
\newcommand{\pairMW}[3]{\pairMX[\weakLabel]{#1}{#2}{#3}}
\newcommand{\pairMS}[3]{\pairMX[\strongLabel]{#1}{#2}{#3}}

\newcommand{\fstName}{\mathsf{fst}}
\newcommand{\fst}[1]{\fstName{} \, #1}
\newcommand{\fstM}[2]{\fstName{}_{#1} \, #2}
\newcommand{\sndName}{\mathsf{snd}}
\newcommand{\snd}[1]{\sndName{} \, #1}
\newcommand{\sndM}[2]{\sndName{}_{#1} \, #2}
\newcommand{\prodrecName}{\mathsf{prodrec}}
\newcommand{\prodrecHead}[2]{\prodrecName{}_{#1}^{#2}}
\newcommand{\prodrec}[5]{\prodrecName{}_{#1}^{#2} \, #3 \, #4 \, #5}
\newcommand{\strongProdrec}[2]{\prodrecName{} \, #1 \, #2}
\newcommand{\prodrecM}[6]{\prodrecName{}_{#1,#2}^{#3} \, #4 \, #5 \, #6}
\renewcommand{\unit}{\star}

\newcommand{\unitS}{\unitX[\strongLabel]}
\newcommand{\unitX}[1][k]{\unit_{#1}}
\newcommand{\unitrecName}{\mathsf{unitrec}}
\newcommand{\unitrec}[5]{\unitrecName_{#1}^{#2} \, #3 \, #4 \, #5}
\newcommand{\emptyrecName}[1][]{\mathsf{emptyrec}_{#1}}
\newcommand{\emptyrec}[3]{\emptyrecName[#1] \, #2 \, #3}

\newcommand{\modeZero}{0_{\textsf{M}}}
\newcommand{\modeOne}{1_{\textsf{M}}}
\newcommand{\modeToGrade}[1]{\overline{#1}}
\newcommand{\gradeToMode}[1]{\underline{#1}}
\newcommand{\modeGradeMode}[2]{#1\odot#2}

\newcommand{\trefl}{\mathsf{refl}}
\newcommand{\refl}[1]{\trefl\,#1}
\newcommand{\tEq}{\mathsf{Eq}}
\newcommand{\Eq}[2]{\tEq\,#1\,#2}

\newcommand{\den}[1]{\llbracket #1 \rrbracket}

\newcommand{\id}{\mathsf{id}}
\newcommand{\stepn}[2]{{\uparrow^{#1}}#2}
\newcommand{\step}[1]{\stepn{}{#1}}
\newcommand{\liftn}[2]{{\Uparrow^{#1}}#2}
\newcommand{\lift}[1]{\liftn{}{#1}}
\newcommand{\wkone}[1]{\subst{#1}{\step{}}}

\newcommand{\subst}[2]{{#1 [#2]}}

\newcommand{\consSubst}[2]{#1, #2}
\newcommand{\substtwo}[3]{\subst{#1}{\consSubst{#3}{#2}}}
\newcommand{\wfSubst}[3]{#1 \vdash #2 \mathrel{:} #3}

\newcommand{\wfSubstM}[2]{#1 \blacktriangleright{} #2}
\newcommand{\substCalc}[2][]{#1\| #2 #1\|}

\newcommand{\emptyCon}{\epsilon}
\newcommand{\snocCon}[2]{#1. #2}
\newcommand{\snocsnocCon}[3]{\snocCon{\snocCon{#1}{#2}}{#3}}
\newcommand{\zeroConM}{\mathbf{0}}

\newcommand{\varConM}[1]{\mathbf{e}_{#1}}
\newcommand{\lookupConM}[2]{#1 (\var{#2})}

\newcommand{\wfVar}[3]{\var{#1} \mathrel{:} #2 \in{} #3}
\newcommand{\wfCon}[1]{\vdash{} #1}
\newcommand{\wfType}[2]{#1 \vdash{} #2}
\newcommand{\wfTerm}[3]{#1 \vdash{} #2 \mathrel{:} #3} %
\newcommand{\eqType}[3]{#1 \vdash{} #2 = #3}
\newcommand{\eqTerm}[4]{#1 \vdash{} #2 = #3 \mathrel{:} #4}
\newcommand{\of}{\!:\!}

\newcommand{\redType}[3]{#1 \vdash{} #2 \longrightarrow{} #3}
\newcommand{\redTerm}[4]{#1 \vdash{} #2 \longrightarrow{} #3 \mathrel{:} #4}
\newcommand{\redsType}[3]{#1 \vdash{} #2 \longrightarrow^* #3}
\newcommand{\redsTerm}[4]{#1 \vdash{} #2 \longrightarrow^* #3 \mathrel{:} #4} %

\newcommand{\redSucTerm}[3][\epsilon]{#1 \vdash{} #2 \longrightarrow^s #3 \mathrel{:} \Nat}
\newcommand{\redsSucTerm}[3][\epsilon]{#1 \vdash{} #2 \longrightarrow^{s*} #3 \mathrel{:} \Nat}

\newcommand{\usage}[2]{#1 \mathrel{\raisebox{1pt}{\ensuremath{\scriptscriptstyle\blacktriangleright}}} #2}
\newcommand{\usageMName}{\mathrel{\raisebox{1pt}{\ensuremath{\scriptscriptstyle\blacktriangleright}}}}
\newcommand{\usageM}[3]{#1 \mathrel{\raisebox{1pt}{\ensuremath{\scriptscriptstyle\blacktriangleright}}}^{#2} #3}
\newcommand{\usageCalc}[2][]{#1| #2 #1|}

\newcommand{\erase}[1]{#1^\bullet}

\newcommand{\eraseType}{\nonstrict{\ensuremath{\loopT}} / \strict{\ensuremath{\undefinedT}}}

\newcommand{\lamT}[1]{\lambda \, #1}
\newcommand{\appT}[2]{#1 \, #2}
\newcommand{\natrecT}[3]{\natrecName{} \, #1 \, #2 \, #3}
\newcommand{\prodrecT}[2]{\prodrecName{} \, #1 \, #2}
\newcommand{\unitrecT}[2]{\unitrecName{} \, #1 \, #2}
\newcommand{\loopT}{\mathord{\circlearrowleft}}
\newcommand{\undefinedT}{\mathord{\lightning}}
\newcommand{\redT}[2]{#1 \longrightarrow{} #2}
\newcommand{\redsT}[2]{#1 \longrightarrow^* #2}
\newcommand{\redSucT}[2]{#1 \longrightarrow^s #2}
\newcommand{\redsSucT}[2]{#1 \longrightarrow^{s*} #2}
\newcommand{\Value}[1]{\mathsf{Value} \, #1}

\definecolor{strictbgr}{RGB}{255,210,210}
\definecolor{strictborder}{RGB}{150,70,70}
\newtcbox{\strict}[1][strictbgr]{%
  on line,
  arc=0pt,
  outer arc=0pt,
  colback=#1,
  colframe=#1,
  boxrule=0.5pt,
  boxsep=0pt,
  left=2pt, right=2pt, top=2pt, bottom=2pt,
  enhanced,
  borderline={0.5pt}{0mm}{strictborder, dash pattern=on 0.5pt off 1pt}
}
\definecolor{nonstrictbgr}{RGB}{200,255,200}
\definecolor{nonstrictborder}{RGB}{70,150,70}
\newtcbox{\nonstrict}[1][nonstrictbgr]{%
  on line,
  arc=0pt,
  outer arc=0pt,
  colback=#1,
  colframe=#1,
  boxrule=0.5pt,
  boxsep=0pt,
  left=2pt, right=2pt, top=2pt, bottom=2pt,
  enhanced,
  borderline={0.5pt}{0mm}{nonstrictborder, dash pattern=on 3pt off 3pt}
}

\newcommand{\isder}[2]{#1 :: #2} 
\newcommand{\rU}{\langle\U\rangle}
\newcommand{\rN}{\langle\Nat\rangle}

\newcommand{\rUnitX}[1]{\langle\UnitX[#1]\rangle}

\newcommand{\rEmpty}{\langle\Empty\rangle}
\newcommand{\rPi}[6]{\langle\Pi_{#1}^{#2}#3#4/#5/#6\rangle}
\newcommand{\rSigma}[6][k]{\langle\Sigma^{#2}_{#1}#3#4/#5/#6\rangle}
\newcommand{\rSigmaX}[6]{\langle\Sigma^{#2}_{#1}#3#4/#5/#6\rangle}

\newcommand{\rNe}[1]{\langle\mathsf{Ne}\,#1\rangle}

\newcommand{\typeCode}[2]{#1 \vDash #2}
\newcommand{\eTerm}[4]{#1 \mathrel{\circledR} #2 \mathrel{:} #3 \mathbin{/} #4}
\newcommand{\eSubst}[4]{#1 \mathrel{\circledR} #2 \mathrel{:} #3 \mathrel{\raisebox{0.5pt}{\ensuremath{\scriptstyle\blacktriangleleft}}} #4}
\newcommand{\eValid}[4]{#2 \mathrel{\raisebox{0.5pt}{\ensuremath{\scriptstyle\blacktriangleright}}} #1 \Vdash #3 \mathrel{:} #4}
\newcommand{\eTermNat}[2]{#1 \mathrel{\circledR_\Nat} #2}

\newcommand{\idfunName}{\mathsf{id}}
\newcommand{\idfunappName}[2]{\app{\app{\idfunName}{0}{#1}}{\omega}{#2}}
\newcommand{\lamN}[3]{\lambda^#1 #2. #3}
\newcommand{\lamTN}[2]{\lambda #1. #2}
\newcommand{\PiTypeN}[5]{\PiType{#1}{#2}{(#3 : #4)}{#5}}
\newcommand{\bindN}[2]{#1.#2}
\newcommand{\snocConN}[3]{#1.#2\mathrel{:}#3}
\newcommand{\sgConN}[2]{#1\mathrel{:}#2}
\newcommand{\idfunN}{\lamN{0}{A}{\lamN{\omega}{x}{x}}}
\newcommand{\idfunappN}[2]{\app{\app{(\idfunN)}{0}{#1}}{\omega}{#2}}

\newcommand{\lSPUNM}[1][\mathcal{M}]{\lambda^{\Sigma\Pi\mathrm{U}\Nat}_{#1}}
\newcommand{\bnfsep}{\;|\;}
\renewcommand{\vec}[1]{\boldsymbol{#1}}

\newcommand{\suck}[2]{\sucName^{#1} \; #2}

\newcommand{\leone}{1^{\text{?}}}

\definecolor{premisebgr}{RGB}{210,210,210}

\newcommand{\prem}[1]{\colorbox{premisebgr}{\ensuremath{#1}}}

\newcommand{\linkFormalization}[3][\defaultModule]{\linkedInline{/#1.html\##2}{#3}}
\newcommand{\linkedInline}[2]{\href{\thmBaseLink#1}{\textcolor{thmLinkColor}{#2}}}
\newcommand{\thmBaseLink}{https://graded-type-theory.github.io/graded-type-theory/jfp-submission-2026-04-30}

\definecolor[named]{ACMDarkBlue}{cmyk}{1,0.58,0,0.21}
\colorlet{thmLinkColor}{ACMDarkBlue}

\newtheorem{theorem}{Theorem}[section]
\newtheorem{definition}{Definition}[section]

\AtEndPreamble{
  \newtheoremstyle{linkedThmStyle}
    {.5\baselineskip plus .2\baselineskip minus .2\baselineskip}
    {.5\baselineskip plus .2\baselineskip minus .2\baselineskip}
    {\itshape}
    {}
    {\scshape}
    {.}
    {.5em}
    {\linkFormalization[\thisModule]{\thisDef}{\thmname{#1}\thmnumber{ #2}\thmnote{ {(#3)}}}} 

  \theoremstyle{linkedThmStyle}
  \newtheorem{inLinkedTheorem}[theorem]{Theorem} 
  \newenvironment{linkedTheorem}[3][]{\def\thisModule{#2}\def\thisDef{#3}\inLinkedTheorem[#1]}{\endinLinkedTheorem}

  \newtheoremstyle{linkedDefStyle}
    {.5\baselineskip plus .2\baselineskip minus .2\baselineskip}
    {.5\baselineskip plus .2\baselineskip minus .2\baselineskip}
    {\normalfont}
    {}
    {\itshape}
    {.}
    {.5em}
    {\linkFormalization[\thisModule]{\thisDef}{\thmname{#1}\thmnumber{ #2}\thmnote{ {(#3)}}}}

  \theoremstyle{linkedDefStyle}
  \newtheorem{inLinkedDefinition}[definition]{Definition}
  \newenvironment{linkedDefinition}[3][]
  {\def\thisModule{#2}\def\thisDef{#3}\inLinkedDefinition[#1]}
  {\endinLinkedDefinition}

}

\theoremstyle{remark}
\newtheorem{remark}{Remark}

\newunicodechar{Σ}{\ensuremath{\mathnormal\Sigma}}
\newunicodechar{ℕ}{\ensuremath{\mathbb{N}}}

\begin{document}

\title{A graded modal dependent type theory with erasure, formalized}

\author{Andreas Abel}
\orcid{0000-0003-0420-4492}
\affiliation{
  \department{Department of Computer Science and Engineering}
  \institution{University of Gothenburg and Chalmers University of Technology}
  \postcode{412~96}
  \city{Göteborg}
  \country{Sweden}
  \authoremail{andreas.abel@gu.se}
}

\author{Nils Anders Danielsson}
\orcid{0000-0001-8688-0333}
\affiliation{
  \department{Department of Computer Science and Engineering}
  \institution{University of Gothenburg and Chalmers University of Technology}
  \postcode{412~96}
  \city{Göteborg}
  \country{Sweden}
  \authoremail{nad@cse.gu.se}
}

\author{Oskar Eriksson}
\orcid{0009-0003-9505-4545}
\affiliation{
  \department{Department of Computer Science and Engineering}
  \institution{University of Gothenburg and Chalmers University of Technology}
  \postcode{412~96}
  \city{Göteborg}
  \country{Sweden}
  \authoremail{oskar.eriksson@gu.se}
}

\begin{abstract}
    We present a graded modal type theory, a dependent type theory with
\emph{grades} that can be used to enforce various properties of the
code.
The theory has $\Pi$-types, weak and strong $\Sigma$-types, natural numbers, an
empty type, and a universe, and we also extend the theory with weak and strong unit
types and graded $\Sigma$-types.
The theory is parameterized by a modality structure, a kind of partially ordered
semiring, whose elements (grades) are used to track the usage of
variables in terms and types.
Different modalities are possible.
We focus mainly on quantitative properties, in particular erasure:
with the erasure modality one can mark function arguments as erasable.

The theory is fully formalized in Agda.
The formalization, which uses a syntactic Kripke logical relation at
its core and is based on earlier work, establishes major
meta-theoretic properties such as consistency,
normalization, and decidability of definitional equality.
We also prove a substitution theorem for grade assignment, and
preservation of grades under reduction.
Furthermore we study an extraction function that translates terms to
an untyped $\lambda$-calculus and removes erasable content, in
particular function arguments with the ``erasable'' grade.
For a certain class of modalities we prove that extraction is sound,
in the sense that programs of natural number type have the same value
before and after extraction.
Soundness of extraction holds also for \emph{open} programs, as
long as all variables in the context are erasable, the context is
consistent, and \emph{erased matches} are not allowed for weak
$\Sigma$ and unit types.
\end{abstract}

\maketitle

\section{Introduction}
Some strongly typed programming languages come with static analyses that determine how variables are used in an expression.
The results of these analyses may contribute to the decision of whether a program is accepted,
or they might inform compiler optimizations or give extra guarantees about accepted programs.
Some of these analyses are:
\begin{enumerate}
\item \emph{Variance:}
  Rocq \citeyearpar{rocq:911} and Agda \citeyearpar{agda:280} have positivity checking for (co)inductive types,
  and Scala has variance checking for higher-order subtyping \citep{DBLP:phd/dnb/Steffen99,DBLP:journals/pacmpl/StuckiG21}.
\item \emph{Irrelevance:}
  Agda allows function arguments to be declared proof-irrelevant \citep{pfenning:intextirr,abelScherer:types10}.
\item \emph{Erasure:}
  Agda and Idris \citep{DBLP:conf/ecoop/Brady21} allow function arguments to be declared irrelevant for execution,
  hence erasable during compilation.
\item \emph{Linearity:}
  In Haskell \citep{DBLP:journals/pacmpl/BernardyBNJS18} and Idris \citep{DBLP:conf/ecoop/Brady21} variables can be declared to be linear.
\end{enumerate}
Both erasure and linearity are supported by \emph{Quantitative Type
  Theory} (QTT) \citep{DBLP:conf/lics/Atkey18}, in which a semiring of
quantities is used to keep track of how variables are used.
QTT is based on earlier work by \citet{DBLP:conf/birthday/McBride16},
and there is related work in simply-typed settings by \citet{DBLP:conf/esop/BrunelGMZ14}, \citet{ghicaSmith:esop14} and \citet{orchard:icfp14}.
It has further been observed that information-flow analyses fit the same framework,
by viewing the lattice of security levels as a partially ordered semiring
\citep{10.1145/2951913.2951939,DBLP:journals/pacmpl/OrchardLE19,DBLP:journals/pacmpl/AbelB20,DBLP:conf/esop/ChoudhuryEW22}.
We use the term \emph{graded modal type theory} \citep{DBLP:conf/esop/MoonEO21} for a type theory with grades taken from a semiring of some kind.

In this work we study a graded modal type theory with full-fledged dependent types, \ie, $\Pi$- and $\Sigma$-types,
a universe, and a type of natural numbers that permits large elimination.
We follow \citet{abel:types18} and \citet{DBLP:conf/esop/MoonEO21} in allowing resources to be tracked in types in addition to in terms (but we do not require such tracking).
The type theory comes with a meta-theory, including soundness of
substitution, normalization, and decidability of
definitional equality, proved via logical relations.
The meta-theory is fully formalized in Agda: we build on an Agda
formalization originally due to \citet{DBLP:journals/pacmpl/0001OV18}.

Our framework supports quantitative instances such as erasure and linearity, and information-flow instances.
However, it does not support instances that affect the definitional equality, like (compile-time) irrelevance.

Different instances of the framework lead to different guarantees for accepted programs.
For example, quantitative instances can enable certain runtime optimizations,
like deallocation of a linear resource after the first access
\citep{DBLP:conf/lics/Atkey18,choudhuryEadesEisenbergWeirich:popl21,eriksson:lic}.
In this article, we focus on instances that allow erasure interpretations and show that actual erasure of parts marked as erasable does not alter program behaviour.
As opposed to syntactic arguments based on reduction relations
\citep{DBLP:conf/fossacs/Mishra-LingerS08,DBLP:conf/birthday/McBride16,DBLP:journals/pacmpl/Tejiscak20},
we employ a logical relation between our type theory and an untyped target language.
This soundness proof is also fully formalized in Agda,
resting on the formalized meta-theory to define the logical relation by recursion on types.

We are not only considering closed programs,
but also \emph{open} programs as long as all their free variables are marked as erasable, and,
importantly, that the open programs live in consistent contexts.
The empty type is inhabited in inconsistent contexts, so in that setting
impossible branches in programs cannot be ruled out.
Several implementations of type theory support the compilation of programs that contain axioms.
For instance, Rocq allows the compilation of programs that contain ``logical axioms'', and the manual warns that ``inconsistent logical axioms may lead to incorrect or non-terminating extracted terms'' \citep{rocq:911}.
We prove formally that it is fine to use consistent axioms, as long as the axioms only appear in erasable positions.

We support two variants of our framework, one with and one without
\emph{erased matches} for weak $\Sigma$-types: if such matches are allowed
then matching on an erased pair is permitted if the obtained
components are only used in erased positions.
Erased matches are akin to Rocq's \emph{singleton elimination} for single-constructor propositions,
and are allowed in some graded modal type theories \citep{DBLP:journals/pacmpl/Tejiscak20,DBLP:conf/esop/MoonEO21,DBLP:journals/pacmpl/AbelB20}
but disallowed in others \citep{DBLP:conf/lics/Atkey18,choudhuryEadesEisenbergWeirich:popl21,DBLP:conf/esop/ChoudhuryEW22}.
Our proof of soundness of extraction does not work for open programs
if erased matches are allowed, but we suspect that soundness could be
proved also in this case (see \Cref{sec:erasedMatching}).

Following \citet{abel:types18} and \citet{DBLP:conf/esop/MoonEO21} we
annotate a $\Pi$-type with \emph{two} grades: one is used for terms of
that type, whereas the other is used for the codomain of the
$\Pi$-type.
We do not make active use of the feature in this text, and our results should hold also for a system without it.
However, perhaps it (or a more restrictive variant in which not all pairs of grades are allowed) could be useful for some type theories where types are
relevant for computation, for instance due to the presence of
\emph{type-case} \citep{DBLP:journals/jfp/CraryWM02}, or for some
variants of cubical type theory
\citep{cohenCoquandHuberMortberg:cubical}.

We make the following contributions:
\begin{enumerate}

\item We extend the ordered semiring of grades with operations and laws used for the recursor for natural numbers
  (\Cref{sec:modalities}), refining previous work that used star-semirings to model recursion \citep{DBLP:conf/esop/BrunelGMZ14}.

\item In \Cref{sec:language} we present a graded modal type theory with large elimination that allows separate usage tracking for terms and types
  \citep{abel:types18,DBLP:conf/esop/MoonEO21}, with a fully formalized meta-theory.
  To the best of our knowledge, this is the first full formalization of the meta-theory
  of a graded dependent type theory with a universe and large elimination.

\item \Cref{sec:usage} contributes a grading judgement for which we prove meta-theoretical properties like substitution and preservation under reduction akin to the work of \citet{DBLP:journals/corr/abs-2005-02247}.  The judgement classifies usage of free variables in expressions and is decidable.

\item In \Cref{sec:erasure} we consider the special case of erasability accounting and give a formalized proof of the correctness of erasure using a logical relation.

\item A discussion of our design choices and alternatives follows in \Cref{sec:discussion},
   in particular concerning the grading of the recursor for natural numbers,
   but also erased matches and unit types.

\item
  In \Cref{sec:modes} we consider an extension with \emph{graded $\Sigma$-types}, similar to types presented by \citet{DBLP:conf/lics/Atkey18} and \citet{choudhuryEadesEisenbergWeirich:popl21},
  that allow us to model record types with erased fields.
  This extension has been formalized as well.

\end{enumerate}

This paper is an extension of our article presented at ICFP 2023 \citep{icfp-version}.
For the reader familiar with this work we highlight the main additions below:
\begin{enumerate}
  \item The natrec-star operator that was previously used to assign grades to the eliminator for natural numbers has been replaced with a new, more flexible operation: ``\nrName{} functions''.
  The motivation behind this change is to avoid issues with expressing \eg{} linearity using natrec-star.
  We discuss this issue further in \Cref{sec:nr} as well as how to define (what we believe to be) suitable \nrName{} functions for some instances.
  \item In our erasure case study (\Cref{sec:erasure}) we now target both a strict (call-by-value) and a non-strict (call-by-name) backend as opposed to only a non-strict one.\footnote{This variant of erasure was previously presented by \citet{danielsson-favier-kubanek-universe-levels}, who build on our work.}
  The latter now uses a more aggressive extraction function that removes both erased lambdas and function arguments as opposed to just replacing arguments with a ``dummy'' term.
  In addition, some dead code is now extracted to a looping term instead of a dummy term to highlight that it is not evaluated.
  \item The technical presentation of the erasure logical relation in \Cref{sec:erasure} is simplified thanks to a new logical relation for types instead of the previous reducibility logical relation.
  \item The weak and strong unit types that were speculatively discussed have now been implemented in the formalization, see \Cref{sec:unit}.
\end{enumerate}
Further, the usage inference and decision procedures that were previously only detailed in an appendix have been incorporated into \Cref{sec:usage}.

Our formalization~\citep{formalization} is a fork of a formalization of decidability of type conversion originally due to \citet{DBLP:journals/pacmpl/0001OV18}.
Definitions and theorems in this paper link to their formalization in an HTML rendering of the \linkFormalization[Everything]{}{Agda code}.
Such links are marked in \linkFormalization[Everything]{}{blue}.

Note that there are differences between the presentation in the
text and the formalization.
For one, there is a single formalization
with parameters that make it possible to control whether the theory
should include things like erased matches for weak $\Sigma$-types
(\Cref{sec:erasedMatching}), unit types
(\Cref{sec:unit}), graded $\Sigma$-types with or without
$\eta$-equality and with particular grades (\Cref{sec:modes}), and one
or two modes (\Cref{sec:modes}).
Another difference is that the formalization uses well-scoped syntax,
whereas the text does not.
In order to aid readability we do not include side conditions related
to things like sizes of contexts in the text.%

In the following section, aimed at domain experts, we locate our work more precisely in the context of contemporary research.
This section may be skipped at first reading, the technical content starts in \Cref{sec:modalities}.

\section{Relation to the state of the art}\label{sec:bigpicture}

Our work is a contribution to an active research field about
modalities and type systems.  We contribute to a corner of the
field where modalities form semirings, or, as a
special case for information-flow analyses, lattices of
security levels.  There are other conceptions of modalities,
such as adjoint intuitionistic linear logics with split contexts \citep{DBLP:conf/lics/CervesatoP96,krishnaswamiPradicBenton:popl15,DBLP:conf/fossacs/Vakar15},
``Fitch-style'' multi-modal type theories \citep{DBLP:journals/lmcs/GratzerKNB21},
where modalities are morphisms
between modes, or (idempotent, monadic) modalities in homotopy type theory \citep{DBLP:journals/lmcs/RijkeSS19}, which are type functors that satisfy certain properties.

In the following we locate our work more precisely in the design/feature
space by presenting and motivating some design decisions.

\subsection{Form of judgement: separate and one-sided usage accounting}

Our central judgement is \fbox{$\usage \gamma t$} where $t$ is an expression
(term or type) and $\gamma$ is a \emph{usage context} assigning each
free variable in $t$ a grade $p$.  We separate this judgement from
the usual typing judgement \fbox{$\wfTerm \Gamma t B$} that expresses that
$t$ has type $B$ in typing context $\Gamma$ which assigns a type to
each variable that is in scope for $t$ and $B$.

Alternatively, we could have considered a single judgement
\fbox{$\wfTerm \Phi t B$} where $\Phi = \gamma\Gamma$ is a context mapping variables to
pairs $pA$ of their type $A$ and their grade $p$.  This is the
dominant conception in the literature
\citep{pfenning:intextirr,reedPierce:icfp10,orchard:icfp14,ghicaSmith:esop14,DBLP:journals/pacmpl/BernardyBNJS18,choudhuryEadesEisenbergWeirich:popl21}.
For simple type systems, this view is equivalent to ours,
since there in the judgement $\wfTerm \Phi t B$
grades only apply to the term $t$,
not to the type $B$ that does not mention any of the variables declared in $\Phi$.

However in dependent type systems, where $\wfTerm \Gamma t B$ usually
implies something like $\wfTerm \Gamma B \Type$, the separation becomes conceptually
important, because $\wfTerm \Phi t B$ may often not imply $\wfTerm
\Phi B \Type$.  It does so
\eg\ in \citeauthor{pfenning:intextirr}'s
modal type theory of intensionality, extensionality and proof
irrelevance \citeyearpar{pfenning:intextirr},
but it does not necessarily do so for quantitative modalities
that denote how often a variable is used in an expression.\footnotemark{}
\footnotetext{For instance, in $\wfTerm{x \of A} {\refl x} {\Eq x x}$,
  the term mentions $x$ once while its type does so twice.}%

\citet{DBLP:conf/birthday/McBride16} pioneered an approach in which
terms $t$ and their types $B$ are considered in two different
worlds: $t$ in a ``material'' world where the resources $\gamma$ for
$t$ are consumed by the construction of $t$, and $B$ in a ``mental''
world where resources do not matter.%
\footnote{McBride uses the words ``consumption'' and ``contemplation'' for what we paraphrased here as ``material'' and ``mental''.}
Consequently, they introduced a
grade $0$ for ``no consumption'' and a judgement we write as
$\wfTerm \Phi t {0B}$.
The latter is essentially bare typing $\wfTerm \Gamma t B$ and
$\Phi$ is there of the form $\vec 0 \Gamma$, \ie{} resource-free.
Their typing rules are formulated via a \emph{two-sided} judgement \fbox{$\wfTerm \Phi t {qB}$},
where both the hypotheses and the conclusion are annotated with grades
drawn from a ``rig'' (a ri\underline{n}g without \underline{n}egation), in particular
$\{0,1,\omega\}$ (``none-one-tons'').\footnotemark{}
\footnotetext{Quantity $0$ for no use (``none''), $1$ for a single use (``one''), and $\omega$ for unrestricted use (``tons'').}
The judgement $\wfTerm {\gamma\Gamma} t {qB}$ was to mean, paraphrased, that ``to produce $q$ copies of the output $t$ we need $\gamma(x)$ copies of each input $(x : A) \in \Gamma$''.
In that sense, grades may also be called \emph{multiplicities}.

The two-sided form $\wfTerm \Phi t {qB}$ has been considered for information flow tracking
\citep{volpano:jcs96,DBLP:conf/esop/ChoudhuryEW22},
where $q$ is the security level of the output and $\Phi$ declares the security levels of the inputs.
However, as \citeauthor{DBLP:conf/lics/Atkey18} observed \citeyearpar{DBLP:conf/lics/Atkey18},
\citeauthor{DBLP:conf/birthday/McBride16}'s calculus fails the substitution property for $q = \omega$ (unrestricted use).
\citeauthor{DBLP:conf/lics/Atkey18} fixed this by giving up the general ``$q$ copies'' intuition
and restricting the right-hand side quantity $q$ to $\{0,1\}$
where the unit $1$ represents the ``material'', resource-aware world and $0$ the ``mental'', purely logical one.

\citeauthor{DBLP:conf/lics/Atkey18}'s formulation is equivalent to splitting the judgement into a one-sided one, $\wfTerm{\gamma\Gamma}tB$ for world $1$, and a non-resourced one, $\wfTerm\Gamma tB$ for world $0$.
From here, we proceed and separate resourcing/usage into its own judgement $\usage \gamma t$.
This works for us because we ensure that $t$ contains enough annotations such that this judgement can be defined without reference to types.
However, note that some of \citeauthor{DBLP:conf/lics/Atkey18}'s term formers may only be used in the non-resourced setting, for instance the projections for his dependent tensor products.
The corresponding $\beta$- and $\eta$-rules are also only present in the non-resourced setting.
Our system does not support this (see also \Cref{sec:refines}).

In \Cref{sec:modes} we consider a variant $\usageM \gamma m t$, which resembles Atkey's two-world accounting.
In that setting we support graded $\Sigma$-types, which are similar to Atkey's dependent tensors.
However, instead of having a single family of types which is ``strong'' in the non-resourced setting and ``weak'' otherwise, we have one variant which is always strong and one which is always weak:
the first projection for the strong graded $\Sigma$-type $\GradedSigmaTypeS{0}{q}{A}{B}$ is only available in the erased setting, but the $\eta$-rule for this type is always active.
We discuss under what conditions $\eta$-expansion is resource-preserving for the strong graded $\Sigma$-types.

\citet{DBLP:journals/lmcs/FuKS22} present a linear dependent type theory for quantum programming with a base type for qubits which is subject to linearity.
Their solution for the coexistence of linearity and dependency is different:
Expressions occurring in types must belong to the ``intuitionistic fragment'' of the language,
in particular not contain qubits or other entities that must adhere to the linearity discipline.
Whenever an expression travels from the term- to the type-side of a judgement, such as in the $\Pi$-elimination rule, its linear parts are erased by the Sh(-) operation.
The intuition is that Sh(-), stripping the qubits, only leaves the \emph{shape} of a value.
Integrating an erasure operation into the typing rules is beyond the capabilities of our system.

\subsection{Usage accounting also in types}
\label{sec:also-in-types}

We take a further step by allowing (but not requiring) an independent usage declaration $\usage \delta B$ for the type $B$ of $t$, rather than ignoring variable usage in $B$.
This has been pioneered by \citet{abel:types18} and \citet{DBLP:conf/esop/MoonEO21}. The latter maintain a triple $(\Delta,\gamma,\delta)$ in the typing judgement $\wfTerm \Gamma t B$, where $\gamma$ and $\delta$ are the resource vectors of $t$ and $B$, respectively, and $\Delta$ is a triangular resource matrix for $\Gamma$, declaring how each hypothesis in $\Gamma$ depends resource-wise on the previous hypotheses.
The full potential of usage accounting in types has not been explored yet, and we do not contribute such applications in this work either.  However, \citeauthor{DBLP:conf/esop/MoonEO21} presented some experiments on optimizing type-checking based on usage annotations in types.

There is another reason why we account for usage also in types:
ignoring usage in types --- or at least in terms that stand for types --- can be unsound in extensions of type theory with aspects of homotopy type theory or cubical type theory (CTT).
\citet{abel-danielsson-vezzosi-erased-univalence} discuss a variant of CTT with support for erasure.
They present examples showing that if the rules for erasure are not set up properly, then one can construct two booleans that are provably distinct but equal after erasure.
The examples do not rely on CTT, it suffices to have an \emph{erased} univalence axiom \citep{hott-book}.

The usage rules that we present for the $\Pi$ and $\Sigma$ type constructors in \Cref{sec:modes} are similar to those presented by \citeauthor{abel-danielsson-vezzosi-erased-univalence}, with one important difference: our type constructors are annotated with \emph{two} grades instead of one.
However, our formalization is parameterized by a \linkFormalization[Definition.Typed.Restrictions]{Type-restrictions.\%CE\%A0\%CE\%A3-allowed}{relation} between those two grades
which in particular makes it possible to force them to be equal.

\subsection{Semantics refines resource-free semantics}
\label{sec:refines}

We maintain the standard definitional equality judgement $\eqTerm \Gamma t {t'} B$ of type theory.
This means that modalities do not interact with equality, limiting the expressiveness of our framework.
For instance, it cannot model proof irrelevance \citep{pfenning:intextirr,abelScherer:types10,DBLP:conf/esop/ChoudhuryEW22,liu:indistinguishability}
or variance \citep{DBLP:phd/dnb/Steffen99,abel:csr06,DBLP:journals/pacmpl/StuckiG21}.
However, this limitation allows us to employ a standard semantics of dependent types and terms,
ignoring grade annotations at first.
The resource-aware semantics is then a \emph{refinement} of that first semantics,
faithful to our separation of the usage relation $\usage \gamma t$ from the typing relation $\wfTerm \Gamma t B$.
Such a two-phase approach has been employed by \citet{DBLP:conf/lics/Atkey18}
who first gives a standard \emph{categories with families} (CwF) semantics to the types and terms of his Quantitative Type Theory (QTT),
and then refines CwFs to \emph{quantitative CwFs} to prove resource-correctness of terms.
In our case, the resource-free semantics is given by a \emph{reducibility} logical relation following \citet{DBLP:journals/pacmpl/0001OV18}, and we refine it for erasure instances by an \emph{erasure} logical relation defined over reducible types.

With simple types, well-typed terms $\wfTerm {x_1 \of p_1A_1, \dots, x_n \of p_1A_n} t B$ in the one-sided judgement
can be interpreted as morphisms
$\den t : D^{p_1}\den{A_1} \times \dots \times D^{p_n}\den{A_n} \to \den B$ for a \emph{graded} comonad $D$ \citep{ghicaSmith:esop14,orchard:icfp14}.
Thus it is natural to refer to the semiring elements $p_i$ as \emph{grades}.
We adopt this terminology even if we do not see how this semantics formally extends to dependent types in the general case.

\subsection{Formalization}

We have fully formalized our results in Agda \citep{agda:280}.
A prime motivation for having mechanized proofs is correctness.
It has already been mentioned that the substitution property failed in the work of \citet{DBLP:conf/birthday/McBride16}.
With a mechanized formalization one can also build on and change previous work without knowing every detail of every proof: if anything breaks, then one is informed of this (barring any bugs in proof checkers).
This has been a big benefit as we have been extending our original work on the formalization \citep{icfp-version}.
Our formalization started out as a fork of the formalization originally due to \citet{DBLP:journals/pacmpl/0001OV18},%
\footnote{Our fork is based on commit \href{https://github.com/mr-ohman/logrel-mltt/tree/bb69ad3f39985b56127ecb303bb625de4787e6c6}{bb69ad3f39}, which includes further developments, some by Oskar Eriksson, Gaëtan Gilbert and Wojciech Nawrocki.}
and our formalization has in turn been the starting point for further work \citep{types25Natrec,eriksson:lic,danielsson-geng-opaque-definitions,danielsson-favier-kubanek-universe-levels}.

\subsection{Universes and large elimination}

Our development includes a universe and a recursor for natural numbers that can compute values and types,
the latter being known as \emph{large elimination}.
Thus we ensure that our results are general enough and scale to full-fledged dependent types.
This generality eliminates common short-cuts in establishing the meta-theory.
For instance, \citet{DBLP:conf/esop/MoonEO21} prove strong normalization via an embedding of reduction into a non-dependent polymorphic language,
adapting the work of \citet{geuvers:shortFlexibleSNCC}.
Such techniques are not available in the presence of large elimination; we instead
pursue the more stony path of dependent logical relations outlined by \citet{DBLP:journals/pacmpl/0001OV18}.
However, this gives us a meta-theory with normalization and decidability of judgemental equality, via a typed \emph{reducibility} logical relation.

\citet{DBLP:conf/lics/Atkey18} presents a more extensional semantics using quantitative CwFs, which also lends itself to large elimination.
This semantics is used to show that linear execution (BCI algebras with some extra structure) forms a model of QTT.
A reviewer of our ICFP 2023 paper \citep{icfp-version} suggested that one could also derive meta-theoretic results like decidability from a suitable CwF, however, as far as we know this has not been spelled out.

\subsection{Instance: erasure, justified by a logical relation}

In \Cref{sec:erasure} we introduce the \emph{erasure} lattice $\omega \leq 0$ (retained vs.\ erased) as a possible instance of our generic graded type theory.
The same instance can also be understood as the 2-point security lattice $\mathsf{L} \leq \mathsf{H}$ (public vs.\ private).

We demonstrate that erasure of $0$-annotated function arguments is sound, \ie{} preserves canonical forms at base type,
via an ``erasure'' logical relation indexed by a logical relation for types.%
\footnote{A reviewer of our ICFP 2023 paper \citep{icfp-version} pointed out that a similar result could perhaps be obtained for QTT by instantiating its realizability semantics \citep[Section~4]{DBLP:conf/lics/Atkey18} in such a way that $!_px$ is a dummy for $p = 0$.}

There is also the syntactic path to correctness of erasure,
by showing that reduction is simulated in the target language
\citep{letouzey:types02,DBLP:conf/fossacs/Mishra-LingerS08,DBLP:conf/birthday/McBride16,DBLP:journals/pacmpl/Tejiscak20}.
The syntactic approach does not rely on normalization or semantics and can thus scale to non-terminating languages.
However, statements like ``if a source term has a value, then the corresponding target term has a value'' may not be enough: it could be the case that a source term does not have a value, and still the corresponding target term has a value.
In the presence of \emph{erased matches} that can be the case for an open term in a consistent, erasable context (see \Cref{sec:erasedMatching}).
We do not solve this problem, but we discuss it.

\citet{choudhuryEadesEisenbergWeirich:popl21} consider a quantitative dependent type theory \textsc{GraD}
(without universes) with a heap semantics.
They prove that quantities correctly predict the number of run-time accesses to heap-stored variables
by giving a reference-counting reduction semantics.
To show that reductions preserve typing, they extend \textsc{GraD} by definitions in the sense of delayed substitutions.
\citet{eriksson:lic} discusses a similar semantics in the context of this formalization where type preservation is proved without using such substitutions.
We do not cover such such reference-counting here, focusing only on erasure.

\subsection{Open terms}
\label{sec:open-terms}

\newcommand{\transform}[2][]{\llbracket#2\rrbracket_{#1}}
\newcommand{\transformName}{\transform{\textunderscore}}
\newcommand{\lttnat}{\Nat}
\newcommand{\lttprop}{\mathit{Prop}}
\newcommand{\ltttm}{\Lambda}

Our proofs of soundness of erasure (\Cref{thm:soundness,thm:soundness-modes}) are stated for \emph{open} terms for which all variables are erasable.

\citet{letouzey-thesis} also supports open terms.
His type theory --- a variant of the Calculus of Inductive Constructions
--- does not use erasure annotations of the kind discussed in this work.
Instead it has a sort $\lttprop$, and terms which have $\lttprop$ as a sort are erased (along with some other things).
Letouzey justifies the correctness of extraction using syntactical logical relation between source and extracted terms implemented as inductive relation in the type theory.
To this end, (open) well-typed terms are translated into proofs of the relation living in a translated version of the typing context of the term.
One question that does not seem to be investigated is whether consistent contexts remain consistent under the translation
--- in the absence of that, extraction correctness would not actually be established in all cases.%
\footnote{More precisely,
\citeauthor{letouzey-thesis} defines a type-preserving transformation $\transformName$ and proves that $\Gamma\vdash t : A$ implies that there is some $u$ such that $\transform[+]{\Gamma}\vdash u : \transform[2]{A}\,\transform{t}\,\mathcal{E}(t)$.
Here $\mathcal{E}(t)$ is the erasure of $t$: the type theory is assumed to contain a type $\ltttm$ that represents untyped terms, and for which the judgemental equality includes some notion of convertibility.
For a type $A$ the translation gives two things: a type $\transform[1]{A}$ and a binary relation $\transform[2]{A}$.
The relation $\transform[2]{\lttnat}$ is an inductively defined predicate of type $\transform[1]{\lttnat}\to\ltttm\to\lttprop$ that holds when either both arguments are zero, or both arguments are applications of successor constructors to related terms.
The context $\transform[+]{\Gamma}$ includes three entries for every binding $x : A$ in $\Gamma$, namely $x : \transform[1]{A}$ and $x' : \ltttm$ and $x'' : \transform[2]{A}\,x\,x'$.
If $A$ is a propositional type, then $\transform[2]{A}\,x\,x'$ always holds, so one obtains a form of correctness of erasure in the presence of propositional postulates.
It does not seem proven that $\transform[1]{A}$ is consistent whenever $A$ is.
Yet to get correctness of extraction in consistent context, the translated contexts need to remain consistent.}

After the publication of our original paper \citep{icfp-version} there has also appeared more work with support for postulates.
\citet{felicissimo-et-al-2026} present an observational type theory with definitionally proof irrelevant propositions and support for propositional postulates.
They prove that, if the postulates are ``validated in set theory'', then every closed term $t$ of type $ℕ$ is observationally equal to a numeral.

\citet{negativeAxioms} discuss how a consistent collection of postulates of the form \mbox{$A\to\Empty$} does not lead to a loss of canonicity for closed terms of type $\Nat$.
Similarly to erased variables, such postulates have no computational content and can thus not be used to produce, \eg, natural numbers.

\section{Modalities organized as ordered semirings}\label{sec:modalities}
We begin by giving the definition of modalities via semiring-like structures.
The definition largely follows that of \citet{DBLP:journals/pacmpl/AbelB20} but is also based on \citeauthor{DBLP:conf/birthday/McBride16}'s rig for erasure and linearity~\citeyearpar{DBLP:conf/birthday/McBride16}, its extension by \citet{DBLP:conf/lics/Atkey18}, and the quantitative coeffect calculus of \citet{DBLP:conf/esop/BrunelGMZ14}, all of which use similar algebraic structures.
\begin{linkedDefinition}[Modality structure]{Graded.Modality}{Modality}\label{def:modality}
  A modality structure is a 7-tuple
  $(\mathcal{M}, +, \cdot, \meet, \nrName, 0, 1)$,\footnote{In our formalization we also include a grade $\omega$ that should satisfy $\omega \le 1$ and $\omega(p + q) \le \omega q$.}
  consisting of (from left to right)
  a type $\mathcal{M}$ of \emph{modalities} or \emph{grades},
  three binary operations (addition, multiplication, and meet),
  a 5-ary function $\nrName$,
  and zero and unit elements,
  satisfying the following properties\footnote{Here, and elsewhere, we occasionally colourize some grades for highlighting.}:
  \begin{itemize}
    \item $(\mathcal{M}, +, \cdot, 0, 1)$ forms a semiring: Addition is commutative and associative with $0$ as identity. Multiplication is associative with $1$ as identity and $0$ as absorbing element and is distributive over addition.
    Like \citet{DBLP:conf/birthday/McBride16}
    we do \emph{not} require $0 \not= 1$, thus,
    the unit type is a valid (but trivial) modality structure.

    As usual, we will typically abbreviate multiplication of $p$ and $q$ as $pq$.

    \item $(\mathcal{M}, \meet)$ forms a semilattice: Meet is commutative, associative, and idempotent. The semilattice induces the usual partial ordering relation: $p \le q$ iff $p = p \meet q$.
    \item Both multiplication and addition distribute over meet.
    \item The function $\nrName$ for the grading of $\bN$-recursion satisfies the following properties (for all grades $p$, $r$, $q_i$, and $q_i'$):%
      \begin{itemize}
        \item Base: If $\qn \le 0$ then $\nr{\ca p}{\cc r}{\qz}{\qs}{\qn} \le \qz$.
        \item Step: $\nr{\ca p}{\cc r}{\qz}{\qs}{\qn} \le \qs + \ca p \cdot \qn + \cc r \cdot \nr{\ca p}{\cc r}{\qz}{\qs}{\qn}$
        \item The function $\nrName$ is monotone in its last three arguments, \ie, if $\qz \le \qzp$, $\qs \le \qsp$, and $\qn \le \qnp$ then $\nr{\ca p}{\cc r}{\qz}{\qs}{\qn} \le \nr{\ca p}{\cc r}{\qzp}{\qsp}{\qnp}$.
        \item Multiplication is right sub-distributive over the last three arguments of $\nrName$, \ie, $(\nr{\ca p}{\cc r}{\qz}{\qs}{\qn}) \cdot \cb q \le \nr{\ca p}{\cc r}{(\qz \cb q)}{(\qs \cb q)}{(\qn \cb q)}$.
        \item Addition is sub-interchangeable with the last three arguments of $\nrName$, \ie, $\nr{\ca p}{\cc r}{\qz}{\qs}{\qn} + \nr{\ca p}{\cc r}{\qzp}{\qsp}{\qnp} \le \nr{\ca p}{\cc r}{(\qz + \qzp)}{(\qs + \qsp)}{(\qn + \qnp)}$.
      \end{itemize}
    \item Equality is decidable.  This property is used in
      decision procedures for typing and usage as well as to differentiate between $0$ and other grades in the erasure case study.
  \end{itemize}
\end{linkedDefinition}

We will typically use $\mathcal{M}$ to refer both to the modality structure and the underlying type.
As we will see, the elements of $\mathcal{M}$ are used to give modal interpretations to programs and, in particular, to free variables.
Which interpretations are available is determined by the specific modality structure that is used.

The operators represent different ways of combining interpretations.
\emph{Addition} combines grades of subterms when both are used, for instance
for components of a weak pair, and \emph{meet} when only one is used, for
instance for components of a strong pair or for branches in a case
distinction.
\emph{Multiplication} allows us to scale by the number of uses.
The operator $\nrName$, which we will refer to as the ``\nrName{} function'' of the modality structure, has a more specific use and will be used to combine grades for use with the natural number eliminator.
We discuss its use, as well as the roles of the five arguments, when we use it to define grade assignment for $\natrecName$ in \Cref{sec:usage}.
The \emph{partial order} encodes a form of compatibility between grades.
Intuitively, $p \leq q$ can be interpreted as allowing $p$ to be used in place of $q$.
Note that from a quantitative perspective, the direction of the order relation is different from what one might first expect since a smaller grade represents a resource that is allowed to be used \emph{more} times.

The properties of modalities are used for our proofs of the grade analogues of subject reduction and the substitution lemma.
For instance, the (sub-) distributivity properties for meet ensure that \linkFormalization[Graded.Modality.Properties.Addition]{+-monotone}{addition} and \linkFormalization[Graded.Modality.Properties.Multiplication]{\%C2\%B7-monotone}{multiplication} are monotone operations.
For the most part, this definition thus matches the one of \citet{DBLP:journals/pacmpl/AbelB20} with the main exception being the \nrName\ function, which is used for $\natrecName$.

Following \citet{orchard:icfp14} and
\citet{DBLP:journals/pacmpl/AbelB20}
we introduce grade or \linkFormalization[Graded.Context]{Con\%E2\%82\%98}{usage contexts}, analogous to typing contexts,
that map free indices of free variables to grades.
We write $\gamma(i)$ for the grade $\gamma$ assigns to index $i$.
We also write $\emptyCon$ for the empty context and $\snocCon{\gamma}{p}$ for the context which assigns grade $p$ to index $0$ and grade $\gamma(i)$ to index $i+1$.
For contexts of the same domain, the operators are \linkFormalization[Graded.Context]{_\%2B\%E1\%B6\%9C_}{lifted to act pointwise} in the following way:
\begin{align*}
    (\gamma + \delta)(i) &= \gamma(i) + \delta(i)  & (p \cdot \gamma)(i) &= p \cdot \gamma(i) \\
    (\gamma \meet \delta)(i) &= \gamma(i) \meet \delta(i)  & (\nr{p}{r}{\gamma}{\delta}{\eta})(i) &= \nr{p}{r}{\gamma(i)}{\delta(i)}{\eta(i)}
\end{align*}
Note that the lifting of $\nrName$ is done only over the last three arguments while the first two remain single grades, and the left argument of multiplication is also a single grade.
Since the operators are defined pointwise, all properties can similarly be lifted to also hold for contexts.
Thus, usage contexts of a given domain \linkFormalization[Graded.Context.Properties]{Con\%E2\%82\%98-semimodule}{form} a \emph{left semimodule} over the semiring, with the zero context, $\zeroConM$, as zero \citep{DBLP:conf/birthday/McBride16}.
The ordering relation is \linkFormalization[Graded.Context]{_\%E2\%89\%A4\%E1\%B6\%9C_}{lifted pointwise} as well, so that $\gamma \leq \delta$ iff $\gamma(x) \leq \delta(x)$ for all $x$ in the domain of $\gamma$ and $\delta$, or equivalently, $\gamma \meet \delta = \gamma$.

The already mentioned \linkFormalization[Graded.Modality.Instances.Unit]{UnitModality}{trivial} (one element) modality structure which reduces the system to the underlying non-graded system is a valid instance.
Below we list some more interesting instances.
For all of them it is possible to define lawful definitions of $\nrName$, however, we mostly defer the discussion of this to \Cref{sec:nr}.

\begin{figure}
  \begin{mathpar}
\textsf{Erasure}\\
\begin{tabular}{l|ll}
$+$      & $0$ & $\omega$ \\ \hline
$0$      & $0$ & $\omega$      \\
$\omega$ & $\omega$ & $\omega$
\end{tabular}
\and
\begin{tabular}{l|ll}
$\cdot$      & $0$ & $\omega$ \\ \hline
$0$      & $0$ & $0$      \\
$\omega$ & $0$ & $\omega$
\end{tabular}
\and
\begin{tabular}{l|ll}
$\meet$      & $0$ & $\omega$ \\ \hline
$0$      & $0$ & $\omega$      \\
$\omega$ & $\omega$ & $\omega$
\end{tabular}\\
\textsf{Affine Types}\\
\begin{tabular}{l|lll}
  $+$      & $0$ & $1$ & $\omega$ \\ \hline
  $0$      & $0$ & $1$ & $\omega$      \\
  $1$      & $1$ & $\omega$ & $\omega$      \\
  $\omega$ & $\omega$ & $\omega$ & $\omega$
\end{tabular}
\and
\begin{tabular}{l|lll}
  $\cdot$      & $0$ & $1$ & $\omega$ \\ \hline
  $0$      & $0$ & $0$ & $0$      \\
  $1$      & $0$ & $1$ & $\omega$      \\
  $\omega$ & $0$ & $\omega$ & $\omega$
\end{tabular}
\and
\begin{tabular}{l|lll}
  $\meet$      & $0$ & $1$ & $\omega$ \\ \hline
  $0$      & $0$ & $1$ & $\omega$      \\
  $1$      & $1$ & $1$ & $\omega$      \\
  $\omega$ & $\omega$ & $\omega$ & $\omega$
\end{tabular}\\
\textsf{Linear Types}\\
\begin{tabular}{l|lll}
  $+$      & $0$ & $1$ & $\omega$ \\ \hline
  $0$      & $0$ & $1$ & $\omega$      \\
  $1$      & $1$ & $\omega$ & $\omega$      \\
  $\omega$ & $\omega$ & $\omega$ & $\omega$
\end{tabular}
\and
\begin{tabular}{l|lll}
  $\cdot$      & $0$ & $1$ & $\omega$ \\ \hline
  $0$      & $0$ & $0$ & $0$      \\
  $1$      & $0$ & $1$ & $\omega$      \\
  $\omega$ & $0$ & $\omega$ & $\omega$
\end{tabular}
\and
\begin{tabular}{l|lll}
  $\meet$      & $0$ & $1$ & $\omega$ \\ \hline
  $0$      & $0$ & $\omega$ & $\omega$      \\
  $1$      & $\omega$ & $1$ & $\omega$      \\
  $\omega$ & $\omega$ & $\omega$ & $\omega$
\end{tabular}\\
\textsf{Linear \& Affine Types}\\
\begin{tabular}{l|llll}
$+$      & $0$ & $1$ & $\leone$ & $\omega$ \\ \hline
$0$      & $0$ & $1$ & $\leone$      & $\omega$      \\
$1$      & $1$ & $\omega$ & $\omega$      & $\omega$      \\
$\leone$ & $\leone$ & $\omega$ & $\omega$      & $\omega$      \\
$\omega$ & $\omega$ & $\omega$ & $\omega$      & $\omega$
\end{tabular}
\and
\begin{tabular}{l|llll}
$\cdot$      & $0$ & $1$ & $\leone$ & $\omega$ \\ \hline
$0$      & $0$ & $0$ & $0$      & $0$      \\
$1$      & $0$ & $1$ & $\leone$      & $\omega$      \\
$\leone$ & $0$ & $\leone$ & $\leone$      & $\omega$      \\
$\omega$ & $0$ & $\omega$ & $\omega$      & $\omega$
\end{tabular}
\and
\begin{tabular}{l|llll}
$\meet$      & $0$ & $1$ & $\leone$ & $\omega$ \\ \hline
$0$      & $0$ & $\leone$ & $\leone$      & $\omega$      \\
$1$      & $\leone$ & $1$ & $\leone$      & $\omega$      \\
$\leone$ & $\leone$ & $\leone$ & $\leone$      & $\omega$      \\
$\omega$ & $\omega$ & $\omega$ & $\omega$      & $\omega$
\end{tabular}
\end{mathpar}
  \Description{Operator definitions for some modality structures.}
  \caption{Operator definitions for some modality structures.}\label{fig:instances}
\end{figure}

\paragraph*{\linkFormalization[Graded.Modality.Instances.Erasure.Modality]{ErasureModality}{Erasure}.}\
The two-element set $\{0, \omega\}$ with $0$ as the additive unit, $\omega$ as the multiplicative unit, operators as defined in \Cref{fig:instances} is a valid modality structure. Note that addition and meet coincide.
The partial order induced by the semilattice is the reflexive closure of $0 \leq \omega$.
For this instance, the \linkFormalization[Graded.Modality.Instances.Erasure.Properties]{nr-unique}{only lawful definition} of $\nrName$ is $\nr{p}{r}{\qzUC}{\qsUC}{\qnUC} = \qzUC \meet \qsUC \meet \qnUC$, independent of $p$ and $r$.
A possible interpretation of this instance is erasure --- the focus of our case study in \Cref{sec:erasure} --- with $0$ denoting computationally irrelevant information and $\omega$ denoting computationally relevant information.
The same instance can also be used for information flow in the lattice $\mathsf{L} \leq \mathsf{H}$
(see \Cref{sec:security}).
\paragraph*{\linkFormalization[Graded.Modality.Instances.Affine]{affineModality}{Affine Types}.}\
The three-element set $\{0, 1, \omega\}$, with $0$ as the additive unit, $1$ as the multiplicative unit and operators as defined in \Cref{fig:instances}, is intended to model affine types.
This interpretation is supported by the partial order $\omega \le 1 \le 0$, with $1$ smaller than $0$: the idea is that the grade $1$ can be interpreted as one or zero uses.
\paragraph*{\linkFormalization[Graded.Modality.Instances.Linearity]{linearityModality}{Linear Types}.}\
The same instance with the meet operator changed in such a way that $1 \ge \omega \le 0$ is intended to model linear types.
Note that the grade $1$ is maximal: the idea is that it stands for exactly one use.
\paragraph*{\linkFormalization[Graded.Modality.Instances.Linear-or-affine]{linear-or-affine}{Linear \& Affine Types}.}\
The set $\{0, 1, \leone, \omega\}$ combines linear and affine types:
$1$ stands for linear uses and $\leone$ for affine uses.
The operators are defined in \Cref{fig:instances} and the partial order is given by $\omega \le \leone$ and $1 \ge \leone \le 0$.
\paragraph*{\linkFormalization[Graded.Modality.Instances.Bounded-distributive-lattice]{modality}{Bounded, Distributive Lattices}.}\
Bounded, distributive lattices with decidable equality can be turned into modality structures where addition is meet and multiplication is join,
the top element is the unit of addition ($0$), and the bottom element is the unit of multiplication ($1$).
A \linkFormalization[Graded.Modality.Instances.Bounded-distributive-lattice]{nr\%E2\%89\%A1\%E2\%88\%A7}{valid \nrName{} function} is given by $\nr{p}{r}{\qzUC}{\qsUC}{\qnUC} = \qzUC \meet \qsUC \meet \qnUC$, the same as for the two-point lattice.
Lattices are used in information flow applications, which we discuss further in \Cref{sec:security}.

\section{Type theory with grades}\label{sec:language}
The language we consider, $\lSPUNM$, is an extension of the language originally used in the formalization by \citet{DBLP:journals/pacmpl/0001OV18}.
Its syntax is shown below.
The language includes the single (Russell) universe $\U$, natural number type $\Nat$, and $\Pi$-type of the original formalization by \citet{DBLP:journals/pacmpl/0001OV18},
as well as the later additions of
  an empty type ($\Empty$) and strong $\Sigma$-types.\footnote{The empty
  type was added to the formalization by Gaëtan Gilbert in 2018. Wojciech Nawrocki added strong $\Sigma$-types and a unit type in 2021. The unit type is discussed in \Cref{sec:unit}.}
This work further adds weak $\Sigma$-types.
The formalization also supports identity types,\footnote{Identity types were added by Nils Anders Danielsson.} a hierarchy of universes and first-class universe levels \citep{danielsson-favier-kubanek-universe-levels} and top-level, possibly opaque definitions \citep{danielsson-geng-opaque-definitions}, but we do not discuss those features.

Since different modality structures $\mathcal{M}$ give different interpretations, $\lSPUNM$ is more accurately seen as a family of languages ranging over the possible structures.

\subsection{Syntax}

Expressions (terms and types) are given by the following grammar:
\begin{align*}
    \linkFormalization[Definition.Untyped]{Term}{A, B, t, u, v ::=}\; &
    \U \bnfsep
    \PiType{p}{q}{A}{B} \bnfsep
    \SigmaTypeS{q}{A}{B} \bnfsep
    \SigmaTypeW{q}{A}{B} \bnfsep
    \Nat\bnfsep
    \Empty
     \\
    \bnfsep & \var{i} \bnfsep \lam{p}{t} \bnfsep \app{t}{p}{u}
    \bnfsep \pairS{t}{u} \bnfsep \fst{t} \bnfsep \snd{t} \bnfsep \pairW{t}{u} \bnfsep \prodrec{r}{q}{A}{t}{u} \\
    \bnfsep & \zero \bnfsep \suc{t} \bnfsep \natrec{p}{q}{r}{A}{t}{u}{v}
    \bnfsep \emptyrec{p}{A}{t}
\TOP{
    \bnfsep \unit
}
\end{align*}
Note that there are two kinds of $\Sigma$-types and pair constructors, annotated with $\strongLabel$ and $\weakLabel$ for strong and weak $\Sigma$-types, respectively.
Often we will not need to differentiate between the two and simply write $\SigmaType{q}{A}{B}$, or $\SigmaTypeX{q}{A}{B}$, and similarly for pairs.
We include two kinds of $\Sigma$-types because the types have different characteristics: the strong kind comes with projections and $\eta$-equality while the weak one comes with pattern matching ($\prodrecName$) and optionally erased matches (and Agda supports all of these variants).

We use de Bruijn-style nameless syntax and write $\var{i}$ for the variable with index $i$.
The subterms which bind new variables are the second argument $B$ of $\Pi$- and $\Sigma$-types, the body $t$ of lambda abstractions, the first argument $A$ of $\natrecName$ and $\prodrecName$ which binds one variable, and the third argument $u$ of $\natrecName$ and $\prodrecName$ which binds two variables.
Since the original formalization \citep{DBLP:journals/pacmpl/0001OV18} the formalized language has been updated to be well-scoped by construction, so subterms are guaranteed to not introduce more variables than specified.

Some terms are annotated with grades, indicated by $p$, $q$, and $r$.
These denote elements from the set of modalities and are used to give programs their desired interpretations based on the chosen instance.
The meaning of an annotation varies somewhat between terms and will be detailed in \Cref{sec:usage}.

\subsection{Weakening and substitution}
\label{sec:weakening-substitution}
\linkFormalization[Definition.Untyped.NotParametrised]{Wk}{Weakenings} $\rho$ are constructed from the identity weaking $\id$ by adding shiftings $\step$ and liftings $\lift$. Identity ($\id$) leaves terms unchanged, shifting ($\step{}$) increases variable indices by one, and lifting ($\lift{}$) is used when a weakening is pushed under a binder.
The application of a weakening $\rho$ to a term $t$ is denoted by \linkFormalization[Definition.Untyped]{wk}{$\subst{t}{\rho}$}.
For variables we have $\subst{\var{i}}{\rho} = \var{\subst{i}{\rho}}$, with the following definition of \linkFormalization[Definition.Untyped.NotParametrised]{wkVar}{$\subst{i}{\rho}$}:
\begin{align*}
  \subst{i}{\id} &= i &
  \subst{i}{\step{\rho}} &= \subst{i}{\rho} + 1 &
  \subst{0}{\lift{\rho}} &= 0 &
  \subst{(i+1)}{\lift{\rho}} &= \subst{i}{\rho} + 1
\end{align*}
For other terms the weakening is applied to each subterm, lifted $n$ times for subterms in which $n$ new variables are bound.

\linkFormalization[Definition.Untyped]{Subst}{Substitutions} are functions from variable indices to terms.
The weakening constructors above can be implemented for substitutions, and we overload the notation.
We also use the notation \linkFormalization[Definition.Untyped]{_\%5B_\%5D}{$\subst{t}{\sigma}$} for the application of a substitution $\sigma$ to a term $t$.
This operation can be implemented similarly to the application of weakenings, with the variable case $\subst{\var{i}}{\sigma} = \sigma(i)$.
Given a substitution $\sigma$ and a term $t$ one can form the \emph{extended} substitution \linkFormalization[Definition.Untyped]{consSubst}{$\consSubst{\sigma}{t}$} satisfying
$(\consSubst{\sigma}{t})(0) = t$ and $(\consSubst{\sigma}{t})(i + 1) = \sigma(i)$.
Note that the substitution $(\consSubst{\id}{u})$ replaces $\var{0}$ with $u$ and reduces the indices of all other free variables by one.
To avoid clutter we will often drop the identity substitution.
For instance, we shall write $\subst{t}{u}$ instead of $\subst{t}{\consSubst{\id}{u}}$
and $\subst t {\step}$ instead of $\subst t {\step\id}$.
Given a substitution $\sigma$ we let \linkFormalization[Definition.Untyped]{head}{$\head{\sigma} = \sigma(0)$} and \linkFormalization[Definition.Untyped]{tail}{$(\tail{\sigma})(i) = \sigma(i + 1)$}.
Note that $\head{(\consSubst{\sigma}{t})} = t$ and $\tail{(\consSubst{\sigma}{t})} = \sigma$.

\subsection{Typing and equality}
\label{sec:typing}
$\lSPUNM$ is dependently typed with the typical judgements:
\linkFormalization[Definition.Typed]{\%E2\%8A\%A2_}{$\wfCon{\Gamma}$} for well-formed contexts,
\linkFormalization[Definition.Typed]{_\%E2\%8A\%A2_}{$\wfType{\Gamma}{A}$} for well-formed types,
\linkFormalization[Definition.Typed]{_\%E2\%8A\%A2_\%E2\%88\%B7_}{$\wfTerm{\Gamma}{t}{A}$} for well-formed terms of type $A$ and \linkFormalization[Definition.Typed]{_\%E2\%8A\%A2_\%E2\%89\%A1_}{$\eqType{\Gamma}{A}{B}$} and
\linkFormalization[Definition.Typed]{_\%E2\%8A\%A2_\%E2\%89\%A1_\%E2\%88\%B7_}{$\eqTerm{\Gamma}{t}{u}{A}$} for type and term equality.
Their inductive definitions are given in \Cref{fig:wf-term,fig:eq-term}.
\linkFormalization[Definition.Untyped.NotParametrised]{Con}{Contexts} are lists of types, each giving a type to a free variable.
We write $\epsilon$ for the empty context and $\snocCon{\Gamma}{A}$ for the context $\Gamma$ extended with $A$.
In that context, $\var{0}$ has type $\wkone{A}$ and $\Gamma$ contains type information for the remaining free variables as expressed by the relation \linkFormalization[Definition.Typed]{_\%E2\%88\%B7_\%E2\%88\%88_}{$\wfVar{i}{B}{\Gamma}$} defined in \Cref{fig:wf-term}.

For the most part these judgements are not affected by the addition of grades.
In order to simplify the meta-theory we include some premises that can be proved from the others.
Premises that can be dropped a posteriori are marked with a grey background.
In some rules certain grades are required to match, for instance in the typing rule for lambda abstractions, where the annotation on the lambda is required to match one of the $\Pi$-type's annotations.
The remaining work of ensuring correctness of grade annotations is handled by the usage relation, which is defined in \Cref{sec:usage}.

\begin{figure}
  \input{Typing.tex}
  \Description{Well-formed contexts, $\wfCon{\Gamma}$, types, $\wfType{\Gamma}{A}$, variables, $\wfVar{i}{A}{\Gamma}$, and terms, $\wfTerm{\Gamma}{t}{A}$.}
  \caption{Well-formed contexts, \linkFormalization[Definition.Typed]{\%E2\%8A\%A2_}{$\wfCon{\Gamma}$}, types, \linkFormalization[Definition.Typed]{_\%E2\%8A\%A2_}{$\wfType{\Gamma}{A}$}, variables, \linkFormalization[Definition.Typed]{_\%E2\%88\%B7_\%E2\%88\%88_}{$\wfVar{i}{A}{\Gamma}$}, and terms, \linkFormalization[Definition.Typed]{_\%E2\%8A\%A2_\%E2\%88\%B7_}{$\wfTerm{\Gamma}{t}{A}$}.}\label{fig:wf-term}
\end{figure}

\begin{figure}[p]
  \input{EqTyping.tex}
  \Description{Equality of types, $\eqType{\Gamma}{A}{B}$, and terms, $\eqTerm{\Gamma}{t}{u}{A}$.}
  \caption{Equality of types, \linkFormalization[Definition.Typed]{_\%E2\%8A\%A2_\%E2\%89\%A1_}{$\eqType{\Gamma}{A}{B}$}, and terms, \linkFormalization[Definition.Typed]{_\%E2\%8A\%A2_\%E2\%89\%A1_\%E2\%88\%B7_}{$\eqTerm{\Gamma}{t}{u}{A}$}.}\label{fig:eq-term}
\end{figure}

One can note the differences between weak and strong $\Sigma$-types.
Some rules are shared between the two variants, while some only apply to one of them.
In particular, the rules involving projections --- including the rule for $\eta$-equality --- are only valid for strong $\Sigma$-types.
Rules involving $\prodrecName$ are only given for weak $\Sigma$-types.

We have typeset certain grade and $\Sigma$-type annotations in colour to highlight some non-standard aspects of the rules.

\begin{remark}[Atkey's dependent tensor product types]
  Our $\Sigma$-types differ from the \emph{dependent tensor product types}
  $(x :^p S) \otimes T$ of \citet{DBLP:conf/lics/Atkey18}.
  In Atkey's case, the grade $p$ is the multiplicity
  of the first component of the \emph{pair},
  whereas in our case this multiplicity is always $1$,
  and the $q$ in $\SigmaTypeX q A B$ is the grade
  of the first component in the \emph{type} $B$.
  Note that \citeauthor{DBLP:conf/lics/Atkey18},
  while including syntax and rules for the dependent tensor product,
  does not spell out its semantics in the given quantitative CwF model.
  We describe a variant of our theory with support for something
  similar to the tensor products of
  \citeauthor{DBLP:conf/lics/Atkey18} --- inspired by his work --- in \Cref{sec:modes}.
\end{remark}

\begin{remark}[Eliminator for $\Sigma_{\strongLabel}$, projections for $\Sigma_{\weakLabel}$]
One can define a variant of $\prodrecName$ for strong $\Sigma$-types by
$\strongProdrec{t}{u} = \subst{u}{\consSubst{\fst{t}}{\snd{t}}}$,
and one can derive a \linkFormalization[Definition.Typed.Properties.Admissible.Sigma]{prodrec\%CB\%A2\%E2\%B1\%BC}{typing rule} for this construction which is similar to the one for weak $\Sigma$-types.
However, this construction does \linkFormalization[Graded.Derived.Sigma]{\%C2\%ACprodrec\%E2\%82\%98}{not} in general satisfy the usage rule that we give for weak $\Sigma$-types in \Cref{sec:usage}.
One can also define projections for weak $\Sigma$-types using $\prodrecName$, but $\eta$-equality does \linkFormalization[Definition.Typed.Consequences.Admissible.Sigma]{\%C2\%AC-\%CE\%A3\%CA\%B7-\%CE\%B7-prod\%CA\%B7-fst\%CA\%B7-snd\%CA\%B7}{not} hold in general for those projections.
\end{remark}

\subsection{Reduction}
\label{sec:reduction}

The operational semantics of $\lSPUNM$ is given by call-by-name weak head reduction relations.
\begin{figure}
  \input{Reductions.tex}%
  \Description{Weak head reduction of types, $\redType{\Gamma}{A}{B}$, and terms, $\redTerm{\Gamma}{t}{u}{A}$, as well as their reflexive, transitive closures $\redsType{\Gamma}{A}{B}$ and $\redsTerm{\Gamma}{t}{u}{A}$.}
  \caption{Weak head reduction of types, \linkFormalization[Definition.Typed]{_\%E2\%8A\%A2_\%E2\%87\%92_}{$\redType{\Gamma}{A}{B}$}, and terms, \linkFormalization[Definition.Typed]{_\%E2\%8A\%A2_\%E2\%87\%92_\%E2\%88\%B7_}{$\redTerm{\Gamma}{t}{u}{A}$}, as well as their reflexive, transitive closures \linkFormalization[Definition.Typed]{_\%E2\%8A\%A2_\%E2\%87\%92\%2A_}{$\redsType{\Gamma}{A}{B}$} and
\linkFormalization[Definition.Typed]{_\%E2\%8A\%A2_\%E2\%87\%92\%2A_\%E2\%88\%B7_}{$\redsTerm{\Gamma}{t}{u}{A}$}.}\label{fig:reduction}
\end{figure}
The reduction relations for types and terms,
\linkFormalization[Definition.Typed]{_\%E2\%8A\%A2_\%E2\%87\%92_}{$\redType{\Gamma}{A}{B}$} and
\linkFormalization[Definition.Typed]{_\%E2\%8A\%A2_\%E2\%87\%92_\%E2\%88\%B7_}{$\redTerm{\Gamma}{t}{u}{A}$} (see \Cref{fig:reduction}),
as well as their reflexive, transitive closures,
\linkFormalization[Definition.Typed]{_\%E2\%8A\%A2_\%E2\%87\%92\%2A_}{$\redsType{\Gamma}{A}{B}$} and
\linkFormalization[Definition.Typed]{_\%E2\%8A\%A2_\%E2\%87\%92\%2A_\%E2\%88\%B7_}{$\redsTerm{\Gamma}{t}{u}{A}$},
are typed, following \citet{types2016decidability,DBLP:journals/pacmpl/0001OV18}.
Terms that are in weak head normal form (WHNF) do not reduce, and reduction to WHNF is deterministic.
\begin{linkedTheorem}[whnfs do not reduce]{Definition.Typed.Properties.Reduction}{whnfRed*Term}
    If\/ $\redsTerm{\Gamma}{t}{u}{A}$ with $t$ in WHNF, then $t = u$.
    Similarly, if\/ $\redsType{\Gamma}{A}{B}$ with $A$ in WHNF, then $A = B$.
\end{linkedTheorem}
\begin{linkedTheorem}[whnf is unique]{Definition.Typed.Properties.Reduction}{whrDet*Term}
    If\/ $\redsTerm{\Gamma}{t}{u}{A}$ and $\redsTerm{\Gamma}{t}{u'}{A'}$ with $u$ and $u'$ in WHNF, then $u = u'$.
    Likewise, if\/ $\redsType{\Gamma}{A}{B}$ and $\redsType{\Gamma}{A}{B'}$ with $B$ and $B'$ in WHNF, then $B = B'$.
\end{linkedTheorem}

\linkFormalization[Definition.Typed.Substitution.Primitive.Primitive]{subst-\%E2\%8A\%A2}{Admissibility of substitution},
\linkFormalization[Definition.Typed.Consequences.Reduction]{whNorm}{normalization}, and
\linkFormalization[Definition.Typed.Decidable.Equality]{decEq}{decidability of equality} are proved either via (well-founded) induction on (sizes of) typing judgements or via
a \emph{reducibility logical relation} adapted from the work of \citet{DBLP:journals/pacmpl/0001OV18}; see the formalization for details.
Furthermore \linkFormalization[Definition.Typed.Decidable]{decConTermType\%E1\%B6\%9C}{type-checking is decidable} for a class of expressions which excludes some redexes.

\section{Assigning grades}\label{sec:usage}
We are now ready to explain the purpose of the different grade annotations and we will do so in parallel with introducing the usage relation which ensures that terms are correctly annotated.
Note that because the usage relation is separate from the typing relation
 --- it is a grade assignment system ---
the logical relation mentioned in the previous section can be defined without reference to usage contexts.
Also note that, unlike for typing, there is only one relation for both terms and types.
Depending on the application, in addition to $\wfTerm{\Gamma}{t}{A}$ and $\usage{\gamma}{t}$ (``$t$ is well-resourced'') it might make sense to require $\usage{\delta}{A}$ (``the type $A$ of $t$ is well-resourced'') and $\usage{\eta_i}{B_i}$ for all $\wfVar{i}{B_i}{\Gamma}$ (``all types in $\Gamma$ are well-resourced'') \citep{DBLP:conf/esop/MoonEO21}.

\begin{linkedDefinition}[Usage relation]{Graded.Usage}{_\%E2\%96\%B8_}\label{def:usage}
    Given a usage context $\gamma$ and term $t$,
    we say that $t$ is \emph{well-resourced} with respect to $\gamma$ if $\usage{\gamma}{t}$.
    That judgement is defined inductively in \Cref{fig:usage}.
\end{linkedDefinition}
\begin{figure}[htbp]
  \input{Usage-def.tex}%
  \Description{The grade assignment relation $\usage{\gamma}{t}$.}
  \caption{The grade assignment relation \linkFormalization[Graded.Usage]{_\%E2\%96\%B8_}{$\usage{\gamma}{t}$}.}\label{fig:usage}
\end{figure}

Similarly to the typing judgements of the previous section, the context is used to keep track of the grades associated to the free variables of a term.
In the explanations to follow, we will stick to a quantitative view where the grades are used to keep track of how many times some resource (term) is referenced or \emph{used}.
It should be noted, however, that the system allows for non-quantitative interpretations as well (see \Cref{sec:security}).

Starting with the simplest cases, the constants, $\U$, $\Nat$, $\Empty$,
and $\zero$, are well-resourced with respect to the zero context $\zeroConM$.
It associates $0$ with every variable in scope, none of which appears in a constant.
A variable $\var{i}$ is well-resourced with respect to the unit-vector context $\varConM{i}$ that maps index $i$ to $1$ and everything else to $0$, representing a single variable occurrence.
These rules are not as restrictive as they may first appear since the subsumption rule (the last rule in the figure) allows us to pass to any pointwise smaller context.
As a reminder, for quantitative modalities a smaller grade represents a resource that is allowed to be used \emph{more} times.

For lambda abstractions $\lam{\ca p}{t}$ the annotation $\ca p$ specifies how many times the newly bound variable (the function argument) may be used in the body $t$.
Through the typing rules discussed in the previous section, this is also the meaning of $\ca p$ in the function type $\PiType{\ca p}{\cb q}{A}{B}$.
The annotation $\cb q$ specifies how many times the function argument is allowed to be used in the function type's codomain $B$.
Analogously, the annotation $\cb q$ on pair types $\SigmaType{\cb q}{A}{B}$ specifies how the first component may be used in the type $B$ of the second component.
For $\Sigma$ and $\Pi$, as well as for weak pairing $\pairW{t}{u}$, the resulting number of uses is given by adding the uses of the subterms.
One can imagine a modality structure where the two annotations $\ca p$ and $\cb q$ on
the $\Pi$-type should not be independent of each other, see \Cref{sec:also-in-types} for one example.
Such cases could be handled by parameterizing typing by a suitable
relation between $\ca p$ and $\cb q$ --- and the formalization has such a parameter.

Applications $\app{t}{\ca p}{u}$ also add the contexts of the two subterms together but with one crucial difference:
the context of the function argument $u$ is first scaled by the annotation $\ca p$, which specifies the number of times the function argument is allowed to be used.
This corresponds to the function argument possibly occurring multiple
times once $\beta$-reduction has been
performed.

In the pairing rule for weak pairs $\pairW{t}{u}$ the contexts of the two components are simply added together as described above.
When such pairs are deconstructed via $\prodrecName$ both components are used, and the usage rule for the pair constructor accounts for both components.

The grade annotation $\ca r$ on $\prodrec{\ca r}{\cb q}{A}{t}{u}$ stands for the number of times the components of the pair can be used, and the context corresponding to the argument $t$ is scaled accordingly.
The annotation $\cb q$ specifies how many times the pair can be used in the type argument.
Note that while we check that the type argument uses resources correctly, its resources are not included in the conclusion of the rule for $\prodrecName$, because type annotations do not contribute to computation.
The case $\ca{r} = 0$ represents matching on the pair without using the components, except possibly in erased positions.
An example is the program $\prodrec{\ca 0}{\cb 0}{\Nat}{\var{0}}{\zero}$ which matches on variable $\var{0}$ and returns $\zero$, roughly corresponding to the following Agda code:
\begin{code}[hide]%
\>[2]\AgdaSymbol{\{-\#}\AgdaSpace{}%
\AgdaKeyword{OPTIONS}\AgdaSpace{}%
\AgdaPragma{--erasure}\AgdaSpace{}%
\AgdaSymbol{\#-\}}\<%
\\
\\[\AgdaEmptyExtraSkip]%
\>[2]\AgdaKeyword{data}\AgdaSpace{}%
\AgdaDatatype{Σ}\AgdaSpace{}%
\AgdaSymbol{(}\AgdaBound{A}\AgdaSpace{}%
\AgdaSymbol{:}\AgdaSpace{}%
\AgdaPrimitive{Set}\AgdaSymbol{)}\AgdaSpace{}%
\AgdaSymbol{(}\AgdaBound{B}\AgdaSpace{}%
\AgdaSymbol{:}\AgdaSpace{}%
\AgdaBound{A}\AgdaSpace{}%
\AgdaSymbol{→}\AgdaSpace{}%
\AgdaPrimitive{Set}\AgdaSymbol{)}\AgdaSpace{}%
\AgdaSymbol{:}\AgdaSpace{}%
\AgdaPrimitive{Set}\AgdaSpace{}%
\AgdaKeyword{where}\<%
\\
\>[2][@{}l@{\AgdaIndent{0}}]%
\>[4]\AgdaOperator{\AgdaInductiveConstructor{\AgdaUnderscore{},\AgdaUnderscore{}}}\AgdaSpace{}%
\AgdaSymbol{:}\AgdaSpace{}%
\AgdaSymbol{(}\AgdaBound{a}\AgdaSpace{}%
\AgdaSymbol{:}\AgdaSpace{}%
\AgdaBound{A}\AgdaSymbol{)}\AgdaSpace{}%
\AgdaSymbol{→}\AgdaSpace{}%
\AgdaBound{B}\AgdaSpace{}%
\AgdaBound{a}\AgdaSpace{}%
\AgdaSymbol{→}\AgdaSpace{}%
\AgdaDatatype{Σ}\AgdaSpace{}%
\AgdaBound{A}\AgdaSpace{}%
\AgdaBound{B}\<%
\\
\\[\AgdaEmptyExtraSkip]%
\>[2]\AgdaKeyword{data}\AgdaSpace{}%
\AgdaDatatype{ℕ}\AgdaSpace{}%
\AgdaSymbol{:}\AgdaSpace{}%
\AgdaPrimitive{Set}\AgdaSpace{}%
\AgdaKeyword{where}\<%
\\
\>[2][@{}l@{\AgdaIndent{0}}]%
\>[4]\AgdaInductiveConstructor{zero}\AgdaSpace{}%
\AgdaSymbol{:}\AgdaSpace{}%
\AgdaDatatype{ℕ}\<%
\\
\>[4]\AgdaInductiveConstructor{suc}\AgdaSpace{}%
\AgdaSymbol{:}\AgdaSpace{}%
\AgdaDatatype{ℕ}\AgdaSpace{}%
\AgdaSymbol{→}\AgdaSpace{}%
\AgdaDatatype{ℕ}\<%
\\
\\[\AgdaEmptyExtraSkip]%
\>[2]\AgdaKeyword{variable}\<%
\\
\>[2][@{}l@{\AgdaIndent{0}}]%
\>[4]\AgdaGeneralizable{A}\AgdaSpace{}%
\AgdaSymbol{:}\AgdaSpace{}%
\AgdaPrimitive{Set}\<%
\\
\>[4]\AgdaGeneralizable{B}\AgdaSpace{}%
\AgdaSymbol{:}\AgdaSpace{}%
\AgdaGeneralizable{A}\AgdaSpace{}%
\AgdaSymbol{→}\AgdaSpace{}%
\AgdaPrimitive{Set}\<%
\end{code}

\begin{code}%
\>[2]\AgdaFunction{ex}\AgdaSpace{}%
\AgdaSymbol{:}\AgdaSpace{}%
\AgdaSymbol{@0}\AgdaSpace{}%
\AgdaDatatype{Σ}\AgdaSpace{}%
\AgdaGeneralizable{A}\AgdaSpace{}%
\AgdaGeneralizable{B}\AgdaSpace{}%
\AgdaSymbol{→}\AgdaSpace{}%
\AgdaDatatype{ℕ}\<%
\\
\>[2]\AgdaFunction{ex}\AgdaSpace{}%
\AgdaSymbol{(\AgdaUnderscore{}}\AgdaSpace{}%
\AgdaOperator{\AgdaInductiveConstructor{,}}\AgdaSpace{}%
\AgdaSymbol{\AgdaUnderscore{})}\AgdaSpace{}%
\AgdaSymbol{=}\AgdaSpace{}%
\AgdaInductiveConstructor{zero}\<%
\end{code}
The annotation \AgdaSymbol{@0} marks the argument of \AgdaFunction{ex} as being erased, corresponding to the marking \eg a lambda or $\prodrecName$ with the grade $\ca{0}$ in our setting.
If $\ca{r} = 0$ (and $0 \neq 1$) then we refer to the match in $\prodrec{\ca r}{\cb q}{A}{t}{u}$ as an \emph{erased match}.
Erased matches are discussed further in \Cref{sec:erasure,sec:erasedMatching}.
The configurable side condition $\Prodrec{\ca r}$ defines which grades $\ca r$ may be used.
We will primarily use it to control whether erased matches are allowed or not.

In contrast to weak pairs, for strong pairs $\pairS{t}{u}$ we instead take the \emph{meet} of the two contexts.
Strong pairs can only be used through projections, and when a projection is used one component is discarded.
The meet of $\cd\gamma$ and $\cf\delta$ is an approximation to both contexts.
If, for instance, a variable occurs linearly in one of the components it will only be treated linearly in the pair if it also occurs linearly in the other component.
If a variable occurs in just one component, it will be treated as if it is used affinely in the pair.
The difference between the two kinds of pairs can also be understood through the lens of linear logic \citep{linearLogic}.
Strong pairs correspond to the multiplicative conjunction $A \otimes B$ where the assumptions $\Gamma,\Delta$ of both conjuncts are concatenated ($\Gamma \otimes \Delta$) when forming the pair,
while weak pairs correspond to the additive conjunction $A \& B$ where both conjuncts should use the same assumptions $\Gamma$.

The eliminator for the empty type, $\emptyrec{\ca p}{A}{t}$, is similar to $\prodrecName$.
However, because uses of this eliminator correspond to dead or impossible branches, we do not restrict the annotation $\ca p$ (apart from the configurable side condition $\Emptyrec{\ca{p}}$), allowing the context to be scaled by any amount.
In particular, $\ca p = 0$ means that we do not count any resources.
A use of this can be found in a \linkFormalization[Graded.Erasure.Examples]{head}{safe \emph{head}} function for (sized) lists, which takes an erasable proof which shows that the list is non-empty, and uses this proof in an application of $\emptyrecName[0]$.
As for $\prodrecName$, we also check the usage in the type argument $A$ and the configurable side condition can be used to control which grades are allowed.

This leaves only $\natrecName$, which
is the only term in our language with recursive and branching semantics, both of which give rise to complications.
The usage relation has primarily been defined with subject reduction in mind.
For terms with (at most) one rule for $\beta$-reduction, doing so is straightforward, but for $\natrecName$ there are two reduction alternatives and we must make sure to include enough resources for either.

This gives rise to two constraints that the context in the conclusion of the rule, call it $\chi$, should satisfy.
In the zero case, $\natrec{\ca p}{\cb q}{\cc r}{A}{z}{s}{\zero}$ reduces to $z$, so for subject reduction to hold we should have $\chi \le \gz$ in this case.
Note, however, that the natural number is then the constant $\zero$ so in this case we have $\gn \le \zeroConM$.

In the successor case, $\natrec{\ca p}{\cb q}{\cc r}{A}{z}{s}{(\suc{n})}$ reduces to $\substtwo{s}{\natrec{\ca p}{\cb q}{\cc r}{A}{z}{s}{n}}{n}$, which uses the resources $\gs + \ca{p} \gn + \cc{r} \chi$ (assuming a suitable substitution lemma), and this gives us the inequality $\chi \le \gs + \ca{p} \gn + \cc{r} \chi$ with $\chi$ on both sides.

If the number of iterations, \ie, the value of $n$, is known, it would be possible to unfold the above definition and arrive at a non-implicit expression for $\chi$, but since this is not always the case we employ a different strategy.
A similar problem has been solved by \citeauthor{DBLP:conf/esop/BrunelGMZ14} who include a fixed-point operator and case distinction on natural numbers in their calculus~\citeyearpar{DBLP:conf/esop/BrunelGMZ14}.
Translated into our setting, their approach roughly corresponds to setting $\chi = \kStar{\cc r}(\gn \meet \gz \meet (\gs + \ca{p} \gn))$, where the unary star-operator satisfies $\kStar{\cc r} = 1 + \cc{r}\kStar{\cc r}$.
Using this rule, however, it seems to be difficult to prove subject reduction in general.%
\footnote{Even if subject reduction would not hold in our system with this rule, this does not mean that the system used by Brunel et al.\ does not have subject reduction. Our form of $\natrecName$ is not definable in their system since their rules for case branching correspond to $\ca{p}$ always being $1$ in our system.}

We have been unable to find a solution to these inequalities that can be expressed using only addition, multiplication and meet for an arbitrary modality.
Instead we include the function $\nrName$ which provides such a solution.
The characteristic inequalities $\nr{\ca p}{\cc r}{\qz}{\qs}{\qn} \le \qz$ when $\qn \le 0$ and $\nr{\ca p}{\cc r}{\qz}{\qs}{\qn} \le \qs + \ca p \qn + \cc r(\nr{\ca p}{\cc r}{\qz}{\qs}{\qn})$ are motivated by the above reasoning, and allow us to prove subject reduction.
In particular, with $\chi = \nr{\ca p}{\cc r}{\gz}{\gs}{\gn}$ we have $\chi \le \gz$ whenever $\gn \le \zeroConM$ and $\chi \le \gs + \ca p\gn + \cc r\chi$ as desired.
The distributivity and interchange properties specify how $\nrName$ relates to the other operators.
These are primarily justified by being used in our proof of the substitution lemma (\Cref{thm:substitution}) and, consequently, subject reduction (\Cref{thm:subject-reduction}).

As a concrete example, consider the erasure modality, for which we noted above that $\nr{\ca p}{\cc r}{\gz}{\gs}{\gn} = \gz \meet \gs \meet \gn$ is the only lawful definition.
This means that a variable is considered to be erased in $\natrec{\ca p}{\cb q}{\cc r}{A}{z}{s}{n}$ only if it is erased in all of $z$, $s$, and $n$, or, phrased differently, that all of $z$, $s$, and $n$ are considered to be computationally relevant.
This should seem reasonable since, with the value of $n$ not being known, both the zero and successor branches may be used at runtime and the natural number argument itself is also relevant simply through matching.
For some other modalities there may be some amount of freedom in how to define $\nrName$.
We will discuss this further in \Cref{sec:nr}.

The usage judgement is algorithmic in the sense that it induces an algorithm for computing a usage context recursively from the term.
This is possible since all variable-binding subterms have a corresponding annotation giving the newly bound variable its grade.
This observation can be formalized via the following function, which infers a usage context from a term.

\begin{linkedDefinition}[Usage inference]{Graded.Usage}{\%E2\%8C\%88_\%E2\%8C\%89}\label{def:usage-inference}
    The usage inference function $\usageCalc{\_}$, which computes a usage context from a term, is defined recursively as follows:
    \begin{align*}
        \usageCalc{\U} & \coloneqq \zeroConM &
        \usageCalc{\zero} & \coloneqq \zeroConM
        \\
        \usageCalc{\Nat} & \coloneqq \zeroConM &
        \usageCalc{\suc{t}} & \coloneqq \usageCalc{t}
        \\
        \usageCalc{\Empty} & \coloneqq \zeroConM &
        \usageCalc{\natrec{\ca p}{\cb q}{\cc r}{A}{z}{s}{n}} & \coloneqq
        \nr{\ca p}{\cc r}{\usageCalc{z}}{(\tail{(\tail{\usageCalc{s}})})}{\usageCalc{n}}
        \\
        \usageCalc{\PiType{\ca p}{q}{A}{B}} & \coloneqq \ca{p}\usageCalc{A} + \tail{\usageCalc{B}} &
        \usageCalc{\pairW{t}{u}} & \coloneqq \usageCalc{t} + \usageCalc{u}
        \\
        \usageCalc{\SigmaType{q}{A}{B}} & \coloneqq \usageCalc{A} + \tail{\usageCalc{B}} &
        \usageCalc{\pairS{t}{u}} & \coloneqq \usageCalc{t} \meet \usageCalc{u}
        \\
        \usageCalc{\var{i}} & \coloneqq \varConM{i} &
        \usageCalc{\fst{t}} & \coloneqq \usageCalc{t}
        \\
        \usageCalc{\lam{p}{t}} & \coloneqq \tail{\usageCalc{t}} &
        \usageCalc{\snd{t}} & \coloneqq \usageCalc{t}
        \\
        \usageCalc{\app{t}{\ca p}{u}} & \coloneqq \usageCalc{t} + \ca{p} \usageCalc{u} &
        \usageCalc{\prodrec{\ca r}{\cb q}{A}{t}{u}} & \coloneqq \ca{r} \usageCalc{t} + \tail{(\tail{\usageCalc{u}})}
        \\
        & &
        \usageCalc{\emptyrec{\ca p}{A}{t}} & \coloneqq \ca{p}\usageCalc{t}
    \end{align*}
\end{linkedDefinition}
\noindent%
Here $(\tail{\gamma})(i) = \gamma(i)$.
Note that $\usageCalc{\_}$ makes no checks on the annotations and will produce a context whether a term is well-resourced or not.
However, if the annotations are correct, then the produced context will not only be valid but will also be an upper bound on valid contexts.
In other words, for any well-resourced term, we can always find the most specific context.
\begin{linkedTheorem}[Principality]{Graded.Usage.Properties}{usage-upper-bound}\label{thm:usageCalc}
    If\/ $\usage{\gamma}{t}$ for some $\gamma$, then $\usage{\usageCalc{t}}{t}$ and $\gamma \le \usageCalc{t}$.
\end{linkedTheorem}
\begin{proof}
    By induction on $\usage{\gamma}{t}$ using the monotonicity of the operators.
\end{proof}
\noindent%
Additionally, we get decidability of the usage relation (if one can decide whether $\Prodrec{r}$ and $\Emptyrec{r}$ hold for any $r$).
\begin{linkedTheorem}[Decidability]{Graded.Usage.Decidable}{\%E2\%8C\%88\%E2\%8C\%89\%E2\%96\%B8\%5B_\%5D\%3F\%E2\%80\%B2_}\label{thm:decUsage}
For any usage context $\gamma$ and term $t$, both $\usage{\usageCalc{t}}{t}$ and $\usage{\gamma}{t}$ are decidable.
\end{linkedTheorem}
\begin{proof}
  A lemma showing that either $\usage{\usageCalc{t}}{t}$, or that $\usage{\gamma}{t}$ does not hold for any $\gamma$, can be proved by induction on $t$.
  Decidability of $\usage{\gamma}{t}$ then follows from decidability of the partial order and \Cref{thm:usageCalc}.
  \end{proof}

  The definition of the usage relation is mainly motivated by a desire to ensure that subject reduction holds. We prove subject reduction using a substitution lemma, and
  the usage relation is designed with that lemma in mind as well.
  When substitutions act on terms they can be considered as maps between terms that may change the number of free variables, or, alternatively, as maps between typing contexts under which the terms are well-formed.
  In the same way, usage substitutions correspond to mappings between usage contexts and these maps turn out to be linear \citep{DBLP:journals/corr/abs-2005-02247,DBLP:journals/pacmpl/AbelB20}, in the sense that they can be \linkFormalization[Graded.Substitution]{Subst\%E2\%82\%98}{represented by matrices} that contain grades.
  By treating usage contexts as vectors for which the $i$-th component is the grade associated to index $i$, substitutions become \linkFormalization[Graded.Substitution]{_\%3C\%2A_}{matrix multiplications}.

  The key to this is, of course, that the matrix must be chosen in such a way that it accurately represents the substitution.
  We generalize the usage relation to substitutions, with $\wfSubstM{\Psi}{\sigma}$ meaning that the substitution matrix $\Psi$ corresponds to the substitution $\sigma$.
  \begin{linkedDefinition}[Usage for substitutions]{Graded.Substitution}{_\%E2\%96\%B6\%5B_\%5D_}
$\wfSubstM{\Psi}{\sigma}$ iff $\forall i. \, \usage{\varConM{i} \Psi}{\sigma(i)}$.
\end{linkedDefinition}
\noindent%
Here $\varConM{i} \Psi$ is the $i$-th row of $\Psi$.
In other words, a valid substitution matrix should have rows corresponding to valid contexts for each term in the substitution.

The identity substitution corresponds to the identity matrix.  A matrix for the single term substitution $(\consSubst{\id}{u})$ --- of the kind that occurs during $\beta$-reduction --- can be constructed by appending a context matching $u$ to the bottom of the identity matrix.

The substitution lemma for grade usage can now be formulated.
\begin{linkedTheorem}[Substitution]{Graded.Substitution.Properties}{subst\%E2\%82\%98-lemma\%E2\%82\%81}\label{thm:substitution}
    If $\usage{\gamma}{t}$ and $\wfSubstM{\Psi}{\sigma}$ then $\usage{\gamma \Psi}{\subst{t}{\sigma}}$.
\end{linkedTheorem}
\begin{proof}
    By induction on $t$ and case analysis for the argument of type $\usage{\gamma}{t}$.
\end{proof}
\noindent%
Note that for $\beta$-reductions, with $\usage{\snocCon{\gamma}{p}}{t}$ and $\usage{\delta}{u}$ and the matrix constructed as above, we get $\usage{\gamma + p \delta}{\subst{t}{u}}$, and $\gamma + p \delta$ is the context that is used for applications in the usage relation.
Correspondences of this kind are used in our proof of subject reduction.
\begin{linkedTheorem}[Subject reduction]{Graded.Reduction}{Subject-reduction.usagePresTerm}\label{thm:subject-reduction}
    If $\usage{\gamma}{t}$ and $\redTerm{\Gamma}{t}{u}{A}$ then $\usage{\gamma}{u}$.
\end{linkedTheorem}
\begin{proof}
    By induction on $\redTerm{\Gamma}{t}{u}{A}$ using \Cref{thm:substitution}.
\end{proof}

Because valid substitution matrices consist of usage contexts corresponding to substituted terms, context inference can be used to infer a substitution matrix from a substitution.
Formally, we define a substitution inference function $\substCalc{\_}$ that computes a substitution matrix from a substitution.

\begin{linkedDefinition}[Substitution usage inference]{Graded.Substitution}{\%E2\%88\%A5_\%E2\%88\%A5}
    The substitution inference function $\substCalc{\_}$, which infers a substitution matrix from a substitution, is defined such that for a substitution taking terms of $m$ free variables to terms of $n$ free variables, $\substCalc{\sigma}$ is an $m \times n$-matrix such that the $i$-th row is $\usageCalc[\big]{\sigma(i)}$.
\end{linkedDefinition}
\noindent%
Similarly to the situation for contexts there is no guarantee that $\substCalc{\sigma}$ is a valid substitution matrix for $\sigma$, but it is valid if a valid substitution matrix exists.
\begin{linkedTheorem}[Principality]{Graded.Substitution.Properties}{subst-calc-correct\%E2\%80\%B2}
    If $\wfSubstM{\Psi}{\sigma}$ for some $\Psi$, then $\wfSubstM{\substCalc{\sigma}}{\sigma}$.
\end{linkedTheorem}
\begin{proof}
    By applying \Cref{thm:usageCalc} to every term in $\sigma$ and its corresponding row in $\Psi$.
\end{proof}

\section{Erasure case study}\label{sec:erasure}
In order to show our system in practice, we now turn our attention to a specific use case of $\lSPUNM$, using a modality for erasure.
The goal is to use annotations to keep track of parts of programs that are not used during evaluation and thus can be safely removed.
We will perform this erasure while compiling programs to an untyped lambda calculus and finally prove  the process to be sound in the sense that it does not affect the outcome of evaluating the program.

We could perform this case study using the erasure modality introduced above.
However, the zero can be interpreted as erasure for other modality structures as well, if the zero satisfies certain properties:

\begin{linkedDefinition}[Well-behaved zero]{Graded.Modality}{Has-well-behaved-zero}\label{def:well-behaved-zero}
    A modality structure is said to have a well-behaved zero if the following properties hold:
    \begin{itemize}
        \item Addition is positive: if $p + q = 0$, then $p = 0$ and $q = 0$.
        \item Meet is positive: if $p \meet q = 0$ then $p = 0$ and $q = 0$.
        \item The \nrName{} function is positive in its last three arguments: if $\nr{p}{r}{\qzUC}{\qsUC}{\qnUC} = 0$ then $\qzUC = 0$, $\qsUC = 0$, and $\qnUC = 0$.
        \item The zero-product property holds: if $pq = 0$ then $p = 0$ or $q = 0$.
        \item The multiplicative zero and unit are distinct, that is, $0 \neq 1$.
    \end{itemize}
\end{linkedDefinition}
The first property is discussed by \citet{DBLP:conf/birthday/McBride16} and the fourth one by \citet{DBLP:conf/lics/Atkey18}.
\citeauthor{DBLP:conf/birthday/McBride16} also discusses the requirement that $0$ should be maximal:\footnote{Since McBride flips the order on grades, our maximal means minimal in their setting.}
this \linkFormalization[Graded.Modality.Properties.Has-well-behaved-zero]{\%F0\%9D\%9F\%98\%E2\%89\%AE}{follows} from the second property.

For the remainder of this section we will work with an arbitrary modality structure with a well-behaved zero.
Examples of such structures include those for \linkFormalization[Graded.Modality.Instances.Erasure.Modality]{erasure-has-well-behaved-zero}{erasure}, \linkFormalization[Graded.Modality.Instances.Linearity]{linearity-has-well-behaved-zero}{linear types}, \linkFormalization[Graded.Modality.Instances.Affine]{affine-has-well-behaved-zero}{affine types} and \linkFormalization[Graded.Modality.Instances.Linear-or-affine]{linear-or-affine-has-well-behaved-zero}{linear or affine types} introduced above, but \linkFormalization[Graded.Modality.Instances.Unit]{\%C2\%ACunit-has-well-behaved-zero}{not the trivial one}.

Though the modality structure may contain any number of elements, for erasure we only need to keep track of whether during evaluation a variable is used or not.
The annotation $0$ will be used to represent the latter and any other annotation will be interpreted as the former.
For convenience, we will borrow the notation of the erasure modality structure and denote any non-zero grade by $\omega$ (if $\omega$ is used in a left-hand side and also in a right-hand side, then both are assumed to refer to the same grade).
Through subsumption, noting that $0 \meet 1 < 0$,
variables that are not used may be labelled as either used or not,
giving the programmer a choice of whether they want them to be erased or not.

We first show that our intended interpretation seems reasonable by the following lemma.
\begin{linkedTheorem}[Variable usage]{Graded.Usage.Properties.Has-well-behaved-zero}{valid-var-usage}\label{thm:eraseVar}
    If $\usage{\gamma}{\var{i}}$ then $\gamma(i) \neq 0$.
\end{linkedTheorem}
\begin{proof}
    From $\usage{\gamma}{\var{i}}$ we can conclude that $\gamma \le \varConM{i}$, and in particular $\gamma(i) \le 1$, but for a well-behaved zero $0 \not\le 1$.
\end{proof}
\noindent%
In short, the lemma states that when the usage relation checks a variable its grade must not be $0$.
This does not necessarily mean, however, that all occurring variables are computationally relevant.
As an example of this, consider the \linkFormalization[Graded.Erasure.Examples]{id}{polymorphic identity function} $\idfunName \coloneq \idfunN$, written here with named variables for convenience, where the type argument $A$ is erased (using the two-element erasure instance for concreteness).
This term has \linkFormalization[Graded.Erasure.Examples]{\%E2\%8A\%A2id}{type} $\PiTypeN{0}{p}{A}{\U}{(\PiTypeN{\omega}{q}{x}{A}{A})}$ and is \linkFormalization[Graded.Erasure.Examples]{\%E2\%96\%B8id}{well-resourced} with respect to the zero context.
In the context $\snocConN{\sgConN{B}{\U}}{b}{B}$ the \linkFormalization[Graded.Erasure.Examples]{id-generic}{term} $\idfunappN{B}{b}$ has \linkFormalization[Graded.Erasure.Examples]{\%E2\%8A\%A2id-generic}{type} $B$.
Since the type argument $B$ is not used during evaluation it should be erasable and indeed, this term is \linkFormalization[Graded.Erasure.Examples]{\%E2\%96\%B8id-generic}{well-resourced} with respect to the context $\snocConN{\sgConN{B}{0}}{b}{\omega}$ (where $b$ is marked as relevant and $B$ as erasable):
\begin{equation*}
    \inferrule*{
        \inferrule*{
            \inferrule*{}
            {
                \usage{\snocConN{\sgConN{B}{0}}{b}{0}}
                      {\idfunN}
            }
            \\
            \inferrule*{ }{\usage{\snocConN{\sgConN{B}{\omega}}{b}{0}}{B}}
        }
        {
            \usage{\snocConN{\sgConN{B}{0}}{b}{0}}
                  {\app{(\idfunN)}{0}{B}}
        }
        \\
        \inferrule*{ }{\usage{\snocConN{\sgConN{B}{0}}{b}{\omega}}{b}}
    }
    {
        \usage{\snocConN{\sgConN{B}{0}}{b}{\omega}}
              {\idfunappN{B}{b}}
    }
\end{equation*}
The key is that for erasable applications, any context can be used to check the function argument since it will anyway be scaled by $0$.
Erasable variables are thus allowed to occur in places such as erasable function arguments.

Since we are now able to track whether a term is computationally relevant or not, we can move on to the task of erasing unneeded parts.
As already mentioned, we do this by compiling $\lSPUNM$ to an untyped lambda calculus.
The syntax of this target language is the following:
\begin{align*}
    \linkFormalization[Graded.Erasure.Target]{Term}{v, w ::=}\;&
    \var{i} \bnfsep \lamT{t} \bnfsep \appT{t}{u}
    \bnfsep \pair{t}{u} \bnfsep \fst{t} \bnfsep \snd{t} \bnfsep \prodrecT{t}{u}
    \bnfsep \zero \bnfsep \suc{t} \bnfsep \natrecT{t}{u}{v}
    \bnfsep \undefinedT
\end{align*}

This language is, except for the lack of types and grades, very similar to $\lSPUNM$ and we re-use the notation for substitutions.
Notably however, the target language also includes the term $\undefinedT$, which represents an undefined value: we treat $\lamT{t}$, $\pair{t}{u}$, $\zero$, $\suc{t}$ and $\undefinedT$ as values.
We support two operational semantics for the target language as shown in \Cref{fig:reductionT}, one \linkFormalization[Graded.Erasure.Target]{_\%E2\%8A\%A2_\%E2\%87\%92_}{non-strict} (call-by-name) in the same style as for $\lSPUNM$, and one \linkFormalization[Graded.Erasure.Target]{_\%E2\%8A\%A2_\%E2\%87\%92_}{strict} (call-by-value).
These differ only in evaluation of applications.
For the strict semantics, the argument must first be reduced to a value before $\beta$-reduction can trigger.
In \Cref{fig:reductionT} we have marked rules and premises which only apply to the strict semantics with a \strict{red} background and we will do so also in the following, similarly marking parts that apply only in the non-strict case with a \nonstrict{green} background.

\begin{figure}[htbp]
    \begin{mathpar}[\lineskip=1.6ex plus 0.3ex]
        \inferrule{\strict{\ensuremath{\Value{w}}}}{\redT{\appT{(\lamT{v})}{w}}{\subst{v}{w}}}
        \and
        \inferrule{\redT{v}{v'}}{\redT{\appT{v}{w}}{\appT{v'}{w}}}
        \and
        \strict{\ensuremath{\inferrule{\redT{w}{w'}\\\Value{v}}{\redT{\appT{v}{w}}{\appT{v}{w'}}}}}
        \\
        \inferrule{\redT{v}{v'}}{\redT{\fst{v}}{\fst{v'}}}
        \and
        \inferrule{\redT{v}{v'}}{\redT{\snd{v}}{\snd{v'}}}
        \and
        \inferrule{ }{\redT{\fst{\pair{v}{w}}}{v}}
        \and
        \inferrule{ }{\redT{\snd{\pair{v}{w}}}{w}}
        \and
        \inferrule{\redT{v}{v'}}{\redT{\prodrecT{v}{w}}{\prodrecT{v'}{w}}}
        \and
        \inferrule{ }{\redT{\prodrecT{\pair{v}{v'}}{w}}{\substtwo{w}{v'}{v}}}
        \and
        \inferrule{\redT{v}{v'}}{\redT{\natrecT{z}{s}{v}}{\natrecT{z}{s}{v'}}}
        \and
        \inferrule{ }{\redT{\natrecT{z}{s}{\zero}}{z}}
        \and
        \inferrule{ }{\redT{\natrecT{z}{s}{(\suc{v})}}{\substtwo{s}{\natrecT{z}{s}{v}}{v}}}
        \and
        \inferrule{ }{\redsT{v}{v}}
        \and
        \inferrule{\redT{v}{v'} \\ \redsT{v'}{w}}{\redsT{v}{w}}
    \end{mathpar}
    \Description{Weak head reduction of target language terms, $\redT{v}{w}$ and its reflexive, transitive closure $\redsT{v}{w}$. Premises and rules with a gray background apply only to the strict semantics.}
    \caption{Weak head reduction of target language terms, \linkFormalization[Graded.Erasure.Target]{_\%E2\%8A\%A2_\%E2\%87\%92_}{$\redT{v}{w}$} and its reflexive, transitive closure \linkFormalization[Graded.Erasure.Target]{_\%E2\%8A\%A2_\%E2\%87\%92\%2A_}{$\redsT{v}{w}$}. Premises and rules with a gray background apply only to the strict semantics.}\label{fig:reductionT}
\end{figure}

The compiler is implemented as a function $\erase{\_}$, taking $\lSPUNM$ terms to terms of the target language.
\begin{linkedDefinition}[Extraction]{Graded.Erasure.Extraction}{erase}
    The extraction function $\erase{\_}$ is defined by recursion on
    terms in \Cref{fig:extraction}.
\end{linkedDefinition}
\begin{figure}[tbp]
  \begin{displaymath}
    \defaultcolumn{@{}l@{}}
    \begin{pmboxed}
      \>\erase{\U} \>\coloneqq \eraseType
      \>[][!@{\qquad\qquad}l@{}]
      \erase{\pair{t}{u}} \>\coloneqq\nonstrict{\ensuremath{\pair{\erase{t}}{\erase{u}}}} /
      \strict{\ensuremath{\appT{\appT{(\lamT{\lamT{\pair{\var{1}}{\var{0}}}})}{\erase{t}}}{\erase{u}}}}\\
      \>\erase{\Nat} \>\coloneqq \eraseType
      \>\erase{(\fst{t})} \>\coloneqq \fst{\erase{t}}\\
      \>\erase{\Empty} \>\coloneqq \eraseType
      \>\erase{(\snd{t})} \>\coloneqq \erase{\snd{t}}\\
      \>\erase{(\PiType{p}{q}{A}{B})} \>\coloneqq \eraseType
      \>\erase{(\prodrec{\ca\omega}{q}{A}{t}{u})} \>\coloneqq \prodrecT{\erase{t}}{\erase{u}}\\
      \>\erase{(\SigmaType{q}{A}{B})} \>\coloneqq \eraseType
      \>\erase{(\prodrec{\ca 0}{q}{A}{t}{u})} \>\coloneqq \substtwo{\erase{u}}{\loopT}{\loopT} \\
      \>\erase{\var{i}} \>\coloneqq \var{i}
      \>\erase{\zero} \>\coloneqq \zero\\
      \>\erase{(\lam{\ca\omega}{t})} \>\coloneqq \lamT{\erase{t}}
      \>\erase{(\suc{t})} \>\coloneqq \nonstrict{\ensuremath{\suc{\erase{t}}}} / \strict{\ensuremath{\appT{(\lamT{(\suc{\var{0}})})}{\erase{t}}}}\\
      \>\erase{(\lam{\ca 0}{t})} \>\coloneqq \nonstrict{\ensuremath{\subst{\erase{t}}{\loopT}}} / \strict{\ensuremath{\lamT{\erase{t}}}}
      \>\erase{(\natrec{p}{q}{r}{A}{z}{s}{n})} \>\coloneqq \natrecT{\erase{z}}{\erase{s}}{\erase{n}} \\
      \>\erase{(\app{t}{\ca\omega}{u})} \>\coloneqq \appT{\erase{t}}{\erase{u}}
      \>\erase{(\emptyrec{p}{A}{t})} \>\coloneqq \loopT\\
      \>\erase{(\app{t}{\ca 0}{u})} \>\coloneqq \nonstrict{\ensuremath{\erase{t}}} / \strict{\ensuremath{\appT{\erase{t}}{\undefinedT}}}
    \end{pmboxed}
  \end{displaymath}
  \Description{Extraction/erasure function. Here $\omega$ stands for any grade distinct from $0$.}
  \caption{\linkFormalization[Graded.Erasure.Extraction]{erase}{Extraction/erasure function}. Here $\omega$ stands for any
  grade distinct from $0$.}\label{fig:extraction}
\end{figure}
\noindent%
Because the languages have similar structure the definition is mostly straightforward, following the structure of the terms.
For type constructors as well as erased lambdas and applications the extraction functions differ depending on if we target the strict or non-strict semantics.

Let us first look at the non-strict case, where there are a couple of things to note.
Firstly, most erasable code is extracted to \linkFormalization[Graded.Erasure.Target.Non-terminating]{loop}{$\loopT$}, which is the \linkFormalization[Graded.Erasure.Target.Non-terminating]{loop-reduces-forever}{looping term} $\appT{(\lamT{\appT{\var{0}}{\var{0}}})}{(\lamT{\appT{\var{0}}{\var{0}}})}$, or $\appT{(\lamTN{x}{\appT{x}{x}})}{(\lamTN{x}{\appT{x}{x}})}$ with named variables.
The idea is that terms which we extract to $\loopT$ should be unused during evaluation, so introducing such loops should not have any bad consequences.
By the end of this section, we will have shown that this is indeed the case.
Applications of type constructors and $\emptyrecName$ are extracted directly to $\loopT$ as these should never be evaluated and have no proper representation in the target language.

Secondly, for lambdas, applications, and $\prodrecName$ we make a case distinction on the grade annotation of the argument or scrutinee, respectively.
When it is non-zero, the extraction procedure preserves the top-level term constructor and extracts the sub-terms (the ``type'' argument is removed for $\prodrecName$).
When the grade is zero, we have encountered an erasable part of the program which is removed by the extraction function.
Erasable function arguments are simply discarded, keeping only the function.
Correspondingly, lambdas taking erasable function arguments are extracted by dropping the lambda and substituting $\loopT$ for the erased argument.
Similarly, for erased uses of $\prodrecName$, the scrutinee is discarded and $\loopT$ is substituted for the components.
As an example, consider \linkFormalization[Graded.Erasure.Examples]{id-\%E2\%84\%95-zero}{$\idfunappName{\Nat}{\zero}$}, the identity function applied to $\Nat$ and $\zero$.
This (closed) term has \linkFormalization[Graded.Erasure.Examples]{\%E2\%8A\%A2id-\%E2\%84\%95-zero}{type} $\Nat$ and is \linkFormalization[Graded.Erasure.Examples]{\%E2\%96\%B8id-\%E2\%84\%95-zero}{well-resourced}.
As we saw above, the type argument is erasable and indeed, both it and the corresponding lambda are removed and we have \linkFormalization[Graded.Erasure.Examples]{erase-non-strict-id-\%E2\%84\%95-zero}{$\erase{(\idfunappN{\Nat}{\zero})} = \appT{(\lamTN{x}{x})}{\zero}$}.

For the most part, extraction to the strict target language is the same but there are a few notable differences.
Since functions are now strict we keep lambdas, and erased function arguments are replaced by the dummy value $\undefinedT$.
Similarly, types are now extracted to $\undefinedT$ instead of $\loopT$.
Note that $\loopT$ is still used in the extraction of $\prodrecName$ and $\emptyrecName$.
In the former case, the components of an erased pair will not be evaluated as before and the latter case still represents an impossible case that will not be evaluated.
We also use lambdas to make pair and successor construction strict.
In this case, the example above is extracted differently, \linkFormalization[Graded.Erasure.Examples]{erase-strict-id-ℕ-zero}{$\erase{(\idfunappN{\Nat}{\zero})} = \appT{\appT{(\lamTN{y}{\lamTN{x}{x}})}{\undefinedT}}{\zero}$}, where both lambdas are kept but the type argument is replaced by $\undefinedT$.

An alternative would have been to keep lambdas, applications and $\prodrecName$ in both the strict and non-strict cases, and instead replace the relevant arguments of applications and $\prodrecName$ by $\undefinedT$ and $\pair{\undefinedT}{\undefinedT}$, respectively.
In our original article \citep{icfp-version} we take such an approach for a non-strict target language.
Conversely, it would \linkFormalization[Graded.Erasure.Consequences.Soundness]{no-run-time-canonicity-if-strict-and-arguments-removed}{not be safe} to remove (all) lambdas in the strict case, since one could then cause a problematic (\eg{} looping or crashing) term to be evaluated.
Our counterexample uses identity types and is similar to one presented by \citet[Figure~5.1]{mishra-linger-2008}.
\citet{letouzey:types02} presents another counterexample that uses an empty type, but that counterexample could seemingly be addressed by defining $\erase{(\emptyrec{p}{A}{t})} \coloneqq \undefinedT$.

If correctly designed, the extraction function should satisfy two main properties: (1) that it erases everything erasable and (2) that it is sound in the sense that it does not affect the outcome of evaluating the program.
The first property is straightforward to prove.

\begin{linkedTheorem}[Erasure is complete]{Graded.Erasure.Extraction.Properties.Usage}{erased-hasX}\label{thm:erasedVar}
    If $\usage{\gamma}{t}$ and $\lookupConM{\gamma}{i} = 0$ then $\var{i}$ does not occur in $\erase{t}$.
\end{linkedTheorem}
\begin{proof}
    By induction on $\usage{\gamma}{t}$ using \Cref{thm:eraseVar}.
\end{proof}

For the example we can see that extraction is sound: the term reduces to $\zero$ both \linkFormalization[Graded.Erasure.Examples]{id-\%E2\%84\%95-zero\%E2\%87\%92\%2Azero}{before} and \linkFormalization[Graded.Erasure.Examples]{erase-id-\%E2\%84\%95-zero\%E2\%87\%92\%2Azero}{after} extraction.
In the remainder of this section we shall demonstrate that the extraction function is sound in general,
meaning that programs retain their values under erasure.

We could restrict the term \emph{programs} to closed expressions of type $\Nat$.
However, in practical dependently typed programming it can be convenient to work under extra assumptions
that are not computationally relevant but can be used to establish correctness properties.
For example, one might want to use the law of excluded middle, a choice principle, some other extension
of type theory, or simply a correctness conjecture that one intends to prove later.
The erasure grades allow us to track which assumptions are not computationally relevant as they only appear in erased parts.
Thus, we aim at correct erasure for programs $\wfTerm {\Delta_0} t \Nat$ with $\usage \zeroConM t$,
meaning that assumptions from $\Delta_0$ appear in $t$ only in erased positions.
This situation is reminiscent of that of \citet{negativeAxioms}, which is discussed in \Cref{sec:open-terms}.

However, in order to support open programs we have to be careful so that $t$ does not get stuck through a match, in a non-erased position, on an erased variable.
In our case, there are two sources of such \emph{erased matches}, \ie{} matches where the formal argument is erased:
\begin{enumerate}
\item
  Eliminating the empty type with $\emptyrecName[0]$.
  We can exclude this either by requiring the side condition $\Emptyrec{0}$ in the usage relation to be false unless $\Delta_0$ is empty, or by requiring our assumptions to be consistent, \ie, there should be no $u$ such that $\wfTerm {\Delta_0} u \Empty$.\footnote{%
  Note that $\lSPUNM$ is normalizing and thus
  \linkFormalization[Definition.Typed.Consequences.Canonicity]{\%C2\%ACEmpty}{the empty context is consistent}.}
\item
  Eliminating a weak pair with $\prodrecHead 0 q$:
  if applied to a neutral pair, $\prodrecName$ does not reduce.
  We disallow matching on erased weak pairs in non-erased positions by requiring the side condition $\Prodrec 0$ to be false unless $\Delta_0$ is empty --- in which case there are no neutral terms to get stuck on.
\end{enumerate}
Some further discussion can be found in \Cref{sec:erasedMatching}.

Our soundness proof uses a logical relation for erasure
which we will simply call \emph{the logical relation}.
The logical relation relates a term $t$ in the source language to a term $v$ in the target language.
It might seem natural to define the relation by induction on the type of $t$, but dependent types do not directly give us a suitable induction principle.
Instead, we first define another logical relation, \emph{the logical relation for types} \linkFormalization[Definition.LogicalRelation.Simplified]{_\%E2\%8A\%A8_}{$\typeCode{\Gamma}{A}$}, which provides us with such a principle.

\begin{linkedDefinition}[Logical relation for types]{Definition.LogicalRelation.Simplified}{_\%E2\%8A\%A8_}
\label{def:reducibility}
    Given a context $\Gamma$, the logical relation for types is expressed by the relation
    $\typeCode{\Gamma}{A}$, defined inductively in \Cref{fig:typeCode}.
    We write $\isder{\mathscr{A}}{J}$ to indicate that $\mathscr{A}$ is a proof of the judgement $J$.
\end{linkedDefinition}
\begin{figure}[htbp]
  \input{Type-code.tex}%
  \Description{The logical relation for types, $\typeCode{\Gamma}{A}$.}
  \caption{The logical relation for types, \linkFormalization[Definition.LogicalRelation.Simplified]{_\%E2\%8A\%A8_}{$\typeCode{\Gamma}{A}$}.}\label{fig:typeCode}
\end{figure}
\noindent%
Roughly speaking, $A$ is in the relation if it reduces to a canonical type with subexpressions in the relation, or if it reduces to a neutral type.
Type families $C$ are in the relation if all of their instantiations $C[t]$ with well-typed terms $t$ are in the relation;
this gives us a suitable induction principle for $\Pi$- and $\Sigma$-types.

The relation is defined inductively with one case for each type former.
We will use labels such as $\rN$
and $\rPi{p}{q}{B}{C}{\mathscr{B}}{\mathscr{C}}$
to indicate the different cases.
In the latter case, $A$ reduces to $\PiType{p}{q}{B}{C}$, the derivation $\mathscr{B}$ shows that $B$ is in the relation, and, for any well-typed term $t$ of type $B$, $\mathscr{C}(t)$ shows that $\subst{C}{t}$ is in the relation.
Note that when we write $\mathscr{C}(t)$, we leave the proof that $t$ is well-typed implicit.

The logical relation for types is based on (a modified version of) the logical relation for reducibility\footnote{See the formalization for details about this relation.} used by \citet{DBLP:journals/pacmpl/0001OV18} but with simplified assumptions, making it easier to work with.
In comparison, the reducibility relation of \citet{DBLP:journals/pacmpl/0001OV18} requires the domain and codomain of $\Pi$- and $\Sigma$-types to be reducible under arbitrary weakenings.
Without such assumptions we do not prove the fundamental lemma for the logical relation for types directly, but we obtain it from the fundamental lemma for the reducibility relation.

\begin{linkedTheorem}[Type reducibility]{Definition.LogicalRelation.Simplified}{\%E2\%8A\%A2\%E2\%86\%92\%E2\%8A\%A8}
  If\/ $\wfType{\Gamma}{A}$ then $\typeCode{\Gamma}{A}$.
\end{linkedTheorem}
\begin{proof}
  By the fundamental lemma for reducibility, $A$ is in the reducibility relation. It follows that $\typeCode{\Gamma}{A}$.
\end{proof}

We will now define the erasure relation $\eTerm{t}{v}{A}{\mathscr{A}}$ in which we relate a source language term $t$ of type $A$ to a target language term $v$.
The relation is defined in such a way that $t$ and $v$ ``behave the same'' when evaluated and the intent is to show that $t$ is related to $\erase{t}$ (\Cref{thm:erasedFundamental}).
We define the relation by recursion on $\mathscr{A}$, a proof that $A$ is in the logical relation for types, essentially allowing us to define it by cases on the type.
As above, there are some differences depending on whether we target the strict or non-strict semantics.
Notably, we add some additional requirements related to strictness when targeting the strict semantics and handle erased functions differently to account for them being extracted differently.

For the remainder of this section, we fix a consistent context $\Delta_0$ with $\wfCon{\Delta_0}$.

\begin{linkedDefinition}[Erasure logical relation]{Graded.Erasure.LogicalRelation}{_\%C2\%AE_\%E2\%88\%B7_\%2F_}
  The logical relation for erasure, $\eTerm{t}{v}{A}{\mathscr{A}}$, where $t$ and $A$ are $\lSPUNM$ terms and $v$ is a target language term, is defined by recursion on $\isder{\mathscr{A}}{\typeCode{\Delta_0}{A}}$ as follows:
  \begin{itemize}
    \item If $\mathscr{A} = \rU$ then $\eTerm{t}{v}{A}{\mathscr{A}}$ holds iff \strict{$\redsT{v}{\undefinedT}$}.
    \item If $\mathscr{A} = \rN$ then $\eTerm{t}{v}{A}{\mathscr{A}}$ iff \linkFormalization[Graded.Erasure.LogicalRelation]{_\%C2\%AE_\%E2\%88\%B7\%E2\%84\%95}{$\eTermNat{t}{v}$}, which is inductively defined to hold iff
    \begin{itemize}
      \item $\redsTerm{\Delta_0}{t}{\zero}{\Nat}$ and $\redsT{v}{\zero}$, or
      \item $\redsTerm{\Delta_0}{t}{\suc{u}}{\Nat}$ and $\redsT{v}{\suc{w}}$ and $\eTermNat{u}{w}$ and \strict{$w$ is a numeral}.
    \end{itemize}
    \item If $\mathscr{A} = \rPi{\ca\omega}{q}{B}{C}{\mathscr{B}}{\mathscr{C}}$ then $\eTerm{t}{v}{A}{\mathscr{A}}$ iff
    \begin{itemize}
      \item $\eTerm{\app{t}{\ca\omega}{u}}{\appT{v}{w}}{\subst{C}{u}}{\mathscr{C}(u)}$ for all $u$ and $w$ with $\wfTerm{\Delta_0}{u}{B}$ and $\eTerm{u}{w}{B}{\mathscr{B}}$, and
      \item \strict{$\redsT{v}{\lamT{v'}}$ for some $v'$.}
    \end{itemize}
    \item If $\mathscr{A} = \rPi{\ca 0}{q}{B}{C}{\mathscr{B}}{\mathscr{C}}$ then $\eTerm{t}{v}{A}{\mathscr{A}}$ iff
    \begin{itemize}
      \item \nonstrict{$\eTerm{\app{t}{\ca 0}{u}}{v}{\subst{C}{u}}{\mathscr{C}(u)}$ for all $u$ with $\wfTerm{\Delta_0}{u}{B}$},
      \item \strict{$\eTerm{\app{t}{\ca 0}{u}}{\appT{v}{\undefinedT}}{\subst{C}{u}}{\mathscr{C}(u)}$ for all $u$ with $\wfTerm{\Delta_0}{u}{B}$}, and
      \item \strict{$\redsT{v}{\lamT{v'}}$ for some $v'$.}
    \end{itemize}
    \item If $\mathscr{A} = \rSigma[\cc k]{\cb q}{B}{C}{\mathscr{B}}{\mathscr{C}}$ then $\eTerm{t}{v}{A}{\mathscr{A}}$ iff there are terms $t_1, t_2, v_1, v_2$ such that
    \begin{itemize}
      \item $\redsTerm{\Delta_0}{t}{\pairX[\cc k]{t_1}{t_2}}{\SigmaTypeX[\cc k]{\cb q}{B}{C}}$,
      \item $\redsT{v}{\pair{v_1}{v_2}}$,
      \item $\wfTerm{\Delta_0}{t_1}{B}$,
      \item $\eTerm{t_1}{v_1}{B}{\mathscr{B}}$, and
      \item $\eTerm{t_2}{v_2}{\subst{C}{t_1}}{\mathscr{C}(t_1)}$.
    \end{itemize}
    \item If $\mathscr{A} = \rEmpty$ or $\mathscr{A} = \rNe{K}$ then $\eTerm{t}{v}{A}{\mathscr{A}}$ does not hold.
  \end{itemize}
\end{linkedDefinition}

For base types, the idea is that the source and target terms should represent the same value of the right type.
In the universe case we put no requirements on either $t$ or $v$ when targeting the non-strict semantics and only require $v$ to reduce to $\undefinedT$ in the strict case.
This is done since $t$ is then a type, and there are no types in the target language.
The natural number case is defined recursively.
Either both terms reduce to $\zero$, or to the successor of some terms which in turn need to be related, and the target term must be a numeral in the strict case.

$\PiName{\ca p}{q}$-types are treated differently depending on whether the function argument is erasable or not.
Non-erasable applications ($\ca p = \omega$) need only be related when the arguments are related.
For erasable functions ($\ca p = 0$) the function arguments do not matter.
In the strict case, we thus use $\undefinedT$ as the argument on the target language side, without requiring it to be related to the source language argument, while in the non-strict case there is no argument at all.
This matches the application cases of the extraction function.
For both erased and non-erased functions we also require the target language term to evaluate to a lambda in the strict case.

$\Sigma$-types are fairly similar to the base types and no difference is made between weak and strong variants:
related terms should reduce to pairs whose components are pairwise related.
Finally, for the empty type and neutral types, no terms are in the logical relation.
The reason for the former is that the consistency of $\Delta_0$ ensures that there are no terms of type $\Empty$.

In order to prove the fundamental lemma we generalize the relation above so that it holds under substitutions to $\Delta_0$.
Let us first define well-typed substitutions.
Intuitively, a substitution $\sigma$ is well-typed, \linkFormalization[Definition.Typed.Substitution.Primitive.Primitive]{_\%E2\%8A\%A2\%CB\%A2_\%E2\%88\%B7_}{$\wfSubst{\Delta}{\sigma}{\Gamma}$}, if all its entries are well-typed terms.
The formal definition is given in \Cref{fig:subst-typing}.

\begin{figure}[h]
  \input{subst-typing.tex}
  \Description{Well-typed substitutions $\wfSubst{\Delta}{\sigma}{\Gamma}$}
  \caption{Well-typed substitutions \linkFormalization[Definition.Typed.Substitution.Primitive.Primitive]{_\%E2\%8A\%A2\%CB\%A2_\%E2\%88\%B7_}{$\wfSubst{\Delta}{\sigma}{\Gamma}$}}\label{fig:subst-typing}
\end{figure}

We then define the logical relation for substitutions, relating substitutions from the source and target languages.

\begin{linkedDefinition}[Substitution logical relation]{Graded.Erasure.LogicalRelation.Hidden}{_\%C2\%AE_\%E2\%88\%B7\%5B_\%5D_\%E2\%97\%82_}
    The logical relation for substitutions,
    $\eSubst{\sigma}{\sigma'}{\Gamma}{\gamma}$,
    relates substitutions $\sigma$ and $\sigma'$ from $\lSPUNM$ and the target language, respectively,
    under contexts $\Gamma$ and $\gamma$.
    It is defined by recursion on the contexts $\Gamma$ and $\gamma$ (which have the same length).
    \begin{itemize}
      \item If $\Gamma = \emptyCon$ and $\gamma = \emptyCon$ then $\eSubst{\sigma}{\sigma'}{\Gamma}{\gamma}$ holds unconditionally.
      \item If $\Gamma = \snocCon{\Delta}{A}$ and $\gamma= \snocCon{\delta}{\ca p}$ then $\eSubst{\sigma}{\sigma'}{\Gamma}{\gamma}$ iff $\eSubst{\tail{\sigma}}{\tail{\sigma'}}{\Delta}{\delta}$ and if $\ca p \neq 0$ then $\eTerm{\head{\sigma}}{\head{\sigma'}}{\subst{A}{\tail{\sigma}}}{\mathscr{A}}$ for some $\isder{\mathscr{A}}{\typeCode{\Delta_0}{\subst{A}{\tail{\sigma}}}}$.
    \end{itemize}
\end{linkedDefinition}
\noindent%
The idea is to think of substitutions as lists of terms, one for each free variable.
These are required to be pairwise related, except for terms corresponding to erasable variables.
The motivation for this is related to \Cref{thm:eraseVar,thm:erasedVar}:
since erasable variables will not occur in non-erasable settings,
and not at all in the target language term,
one should be able to substitute anything (of the correct type) for them without consequence.

The final piece of the logical relation is erasure validity, which both expresses the closure of the logical relation under substitutions and our goal that terms should be related to their extracted counterparts.
\begin{linkedDefinition}[Validity]{Graded.Erasure.LogicalRelation.Hidden}{_\%E2\%96\%B8_\%E2\%8A\%A9\%CA\%B3_\%E2\%88\%B7\%5B_\%5D_}
  The validity relation for erasure,
  $\eValid{\Gamma}{\gamma}{t}{A}$,
  is defined to hold iff for all substitutions $\sigma$, $\sigma'$
  such that $\wfSubst{\Delta_0}{\sigma}{\Gamma}$
  and $\eSubst{\sigma}{\sigma'}{\Gamma}{\gamma}$
  we have $\eTerm{\subst{t}{\sigma}}{\subst{\erase{t}}{\sigma'}}{\subst{A}{\sigma}}{\mathscr{A}}$
  for some $\isder{\mathscr{A}}{\typeCode{\Delta_0}{\subst{A}{\sigma}}}$.
\end{linkedDefinition}

Before proving the fundamental lemma we prove two other lemmas.
The first is closure under expansion for both source and target terms.

\begin{linkedTheorem}[Closure under expansion]{Graded.Erasure.LogicalRelation.Reduction}{redSubstTerm*}\label{thm:LRred}
  If $\isder{\mathscr{A}}{\typeCode{\Gamma}{A}}$
  and $\redsTerm{\Delta_0}{t}{t'}{A}$ and $\redsT{v}{v'}$ and $\eTerm{t'}{v'}{A}{\mathscr{A}}$ then $\eTerm{t}{v}{A}{\mathscr{A}}$.
\end{linkedTheorem}
\begin{proof}
  By induction on $\eTerm{t'}{v'}{A}{\mathscr{A}}$.
\end{proof}

The second property can be seen as a different form of the subsumption rule from the usage relation, allowing us to change the quantity or grade contexts in the substitution and validity relations.
\begin{linkedTheorem}[Subsumption]{Graded.Erasure.LogicalRelation.Hidden}{subsumption-\%C2\%AE\%E2\%88\%B7\%5B\%E2\%88\%A3\%5D\%E2\%97\%82}\label{thm:subsumption}
    If $\gamma(i)=0$ implies $\delta(i)=0$ for all indices $i$, then
  \begin{itemize}
      \item if $\eSubst{\sigma}{\sigma'}{\Gamma}{\gamma}$, then $\eSubst{\sigma}{\sigma'}{\Gamma}{\delta}$, and
      \item if $\eValid{\Gamma}{\delta}{t}{A}$ then $\eValid{\Gamma}{\gamma}{t}{A}$.
  \end{itemize}
\end{linkedTheorem}
\begin{proof}
    For substitutions, by induction on $\Gamma$ and $\gamma$.
    The validity case then follows.
\end{proof}
\noindent%
This property is important for our proof of the fundamental lemma since it allows us to re-fit a proof that substitutions are related to the correct usage context for each subterm.

\begin{linkedTheorem}[Fundamental lemma]{Graded.Erasure.LogicalRelation.Fundamental}{Fundamental.fundamental}\label{thm:fundamental}
    If\/ $\Prodrec{0}$ is false or $\Delta$ is empty,
    $\Delta$ is consistent if $\Emptyrec{0}$ is true,
    and furthermore $\wfTerm{\Gamma}{t}{A}$ and $\usage{\gamma}{t}$, then $\eValid{\Gamma}{\gamma}{t}{A}$.
\end{linkedTheorem}
\begin{proof}
    By induction on $\wfTerm{\Gamma}{t}{A}$ using \Cref{thm:LRred,thm:subsumption,thm:eraseVar}.
\end{proof}

The fundamental lemma implies that for any related substitutions $\sigma$ and $\sigma'$, $\subst{t}{\sigma}$ is related to $\subst{\erase{t}}{\sigma'}$.
We would like to apply this result to identity substitutions, but naturally first need to show that these are related.
More generally, \emph{all} substitutions \emph{from erasable contexts} are related.

\begin{linkedTheorem}[Erased substitutions]{Graded.Erasure.LogicalRelation.Hidden}{\%C2\%AE\%E2\%88\%B7\%5B\%5D\%E2\%97\%82\%F0\%9D\%9F\%98\%E1\%B6\%9C}\label{thm:erasedSubst}
  $\eSubst{\sigma}{\sigma'}{\Gamma}{\zeroConM}$
  for all $\sigma$ and $\sigma'$.
\end{linkedTheorem}

\begin{linkedTheorem}[Logical erasure soundness]{Graded.Erasure.LogicalRelation.Fundamental}{Fundamental.fundamentalErased-\%F0\%9D\%9F\%99\%E1\%B5\%90}\label{thm:erasedFundamental}
  If\/ $\Prodrec{0}$ is false or $\Delta$ is empty,
  $\Delta$ is consistent if $\Emptyrec{0}$ is true, and furthermore $\wfTerm{\Delta_0}{t}{A}$ and $\usage{\zeroConM}{t}$ then there is $\isder{\mathscr{A}}{\typeCode{\Delta_0}{A}}$ such that $\eTerm{t}{\erase{t}}{A}{\mathscr{A}}$.
\end{linkedTheorem}
\begin{proof}
  By the fundamental lemma applied to identity substitutions and \Cref{thm:erasedSubst}.
\end{proof}

With the fundamental lemma proven, we can finally move on to showing that the extraction function is sound for programs computing natural numbers.
In the source language, terms only reduce to weak head normal form so the final result of evaluating such a program will either be $\zero$ or $\suc{t}$ for some $t$ which might not be in WHNF.
The same is true for the target language when the non-strict semantics is used (in the strict case the logical relation requires that $t$ is a numeral).
In other words, reducing terms to WHNF is not enough to determine whether two terms represent the same number or not.
To circumvent this we extend the reduction relations of both the source and the non-strict target languages to allow reduction under $\sucName$, see \Cref{fig:sucred},
with reflexive-transitive closures $\longrightarrow^{s*}$ as expected.
\begin{figure}
\begin{mathpar}
    \inferrule{\redTerm{\Gamma}{t}{t'}{\Nat}}{\redSucTerm[\Gamma]{t}{t'}}
    \and
    \inferrule{\redSucTerm[\Gamma]{t}{t'}}{\redSucTerm[\Gamma]{\suc{t}}{\suc{t'}}}
    \and
    \inferrule{\redT{v}{v'}}{\redSucT{v}{v'}}
    \and
    \nonstrict{\ensuremath{\inferrule{\redSucT{v}{v'}}{\redSucT{\suc{v}}{\suc{v'}}}}}
\end{mathpar}
\Description{Extended reduction relations, $\redSucTerm[\Gamma]{t}{u}$ and $\redSucT{t}{u}$.}
\caption{Extended reduction relations, \linkFormalization[Graded.Erasure.SucRed]{_\%E2\%8A\%A2_\%E2\%87\%92\%CB\%A2_\%E2\%88\%B7\%E2\%84\%95}{$\redSucTerm[\Gamma]{t}{u}$} and \linkFormalization[Graded.Erasure.SucRed]{_\%E2\%8A\%A2_\%E2\%87\%92\%CB\%A2_}{$\redSucT{t}{u}$}.}\label{fig:sucred}
\end{figure}

Writing $\numeral{n}$ for the numeral $\suck n \zero$, where $\suck{0}{t}$ is $t$ and $\suck{k+1}{t}$ is $\suc{(\suck{k}{t})}$, we give the soundness theorem.
\begin{linkedTheorem}[Operational erasure soundness]{Graded.Erasure.Consequences.Soundness}{Soundness\%E2\%80\%B2.Soundness.soundness-\%E2\%84\%95}\label{thm:soundness}
    Let $\Delta$ be a context.
    If\/ $\Prodrec{0}$ is false or $\Delta$ is empty,
    $\Delta$ is consistent if $\Emptyrec{0}$ is true,
    and furthermore
    $\wfTerm{\Delta}{t}{\Nat}$ and $\usage{\zeroConM}{t}$, then
    there is a natural number $n$ such that
    $\redsSucTerm[\Delta]{t}{\numeral{n}}$ and $\redsSucT{\erase{t}}{\numeral{n}}$.
\end{linkedTheorem}
\begin{proof}
  By \Cref{thm:erasedFundamental} we have $\eTerm{t}{\erase{t}}{\Nat}{\rN}$.
  The proof proceeds by induction on this derivation.
\end{proof}

\section{Discussion}\label{sec:discussion}
In total the formalization, including the erasure case study, the extensions discussed in \Cref{sec:unit,sec:modes}, and various other additions,
consists of roughly \num{170000} lines (\SI{7900}{\kilo\byte})
of Agda code, comments and empty lines not included.
This can be compared to the roughly \num{13000} lines of code (\SI{700}{\kilo\byte})
forked from an extension of the development of \citet{DBLP:journals/pacmpl/0001OV18}.
However, the current formalization contains substantial extensions not discussed in this paper, such as identity types, universe polymorphism, further modality structures, and a resource-aware abstract machine to verify quantitative typing beyond erasure.
These extensions contribute roughly half of the total lines of code.

While some time was spent modifying the Agda development, we believe that extending a pre-existing formalization overall provided great benefit, as it simplified the development process in several ways.
For instance, we did not have to define the language from scratch and the code already included proofs of many non-trivial properties.
We also found the reducibility relation useful as the starting point for defining the logical relation in the erasure case study.

\subsection{Usage counting for the natural number eliminator}\label{sec:nr}

Although the erasure modality structure only has \linkFormalization[Graded.Modality.Instances.Erasure.Properties]{nr-unique}{one possible choice} for its \nrName{} function, this is not the case in general.
This leads to some care being required to ensure that one achieves the intended interpretation of the modality.
Taking the linear types structure as an example, one can define \linkFormalization[Graded.Modality.Instances.Zero-one-many]{zero-one-many-greatest-star-nr}{a lawful \nrName{} function} which fails to encode linearity.
The issue can be seen by considering a definition of \linkFormalization[Definition.Untyped.Nat]{plus\%E2\%80\%B2}{addition} for natural numbers,
 $\natrec{0}{0}{1}{(\bindN{m}{\Nat})}{k}{(\bindN{m}{\bindN{r}{\suc{r}}})}{n}$,
 adding variables $k$ and $n$.
It corresponds to the following Agda code which would appear to be linear in both its arguments --- corresponding definitions are accepted as linear by both Idris~2 and Linear Haskell:
\begin{code}[hide]%
\>[2]\AgdaKeyword{data}\AgdaSpace{}%
\AgdaDatatype{ℕ}\AgdaSpace{}%
\AgdaSymbol{:}\AgdaSpace{}%
\AgdaPrimitive{Set}\AgdaSpace{}%
\AgdaKeyword{where}\<%
\\
\>[2][@{}l@{\AgdaIndent{0}}]%
\>[4]\AgdaInductiveConstructor{zero}\AgdaSpace{}%
\AgdaSymbol{:}\AgdaSpace{}%
\AgdaDatatype{ℕ}\<%
\\
\>[4]\AgdaInductiveConstructor{suc}\AgdaSpace{}%
\AgdaSymbol{:}\AgdaSpace{}%
\AgdaDatatype{ℕ}\AgdaSpace{}%
\AgdaSymbol{→}\AgdaSpace{}%
\AgdaDatatype{ℕ}\<%
\end{code}

\begin{code}%
\>[2]\AgdaFunction{plus}\AgdaSpace{}%
\AgdaSymbol{:}\AgdaSpace{}%
\AgdaDatatype{ℕ}\AgdaSpace{}%
\AgdaSymbol{→}\AgdaSpace{}%
\AgdaDatatype{ℕ}\AgdaSpace{}%
\AgdaSymbol{→}\AgdaSpace{}%
\AgdaDatatype{ℕ}\<%
\\
\>[2]\AgdaFunction{plus}\AgdaSpace{}%
\AgdaBound{k}\AgdaSpace{}%
\AgdaInductiveConstructor{zero}%
\>[18]\AgdaSymbol{=}\AgdaSpace{}%
\AgdaBound{k}\<%
\\
\>[2]\AgdaFunction{plus}\AgdaSpace{}%
\AgdaBound{k}\AgdaSpace{}%
\AgdaSymbol{(}\AgdaInductiveConstructor{suc}\AgdaSpace{}%
\AgdaBound{n}\AgdaSymbol{)}\mbox{ }%
\>[18]\AgdaSymbol{=}\AgdaSpace{}%
\AgdaInductiveConstructor{suc}\AgdaSpace{}%
\AgdaSymbol{(}\AgdaFunction{plus}\AgdaSpace{}%
\AgdaBound{k}\AgdaSpace{}%
\AgdaBound{n}\AgdaSymbol{)}\<%
\end{code}
However, with the \nrName{} function linked above, both arguments are instead \linkFormalization[Graded.Modality.Instances.Linearity.Examples.Bad.Nr]{\%E2\%96\%B8plus\%E2\%80\%B2-x\%E2\%82\%80-x\%E2\%82\%81}{assigned grade $\omega$}, representing a non-linear usage.
Worse, when adding a variable to itself, the variable would be considered to be \linkFormalization[Graded.Modality.Instances.Linearity.Examples.Bad.Nr]{\%E2\%96\%B8plus\%E2\%80\%B2-x\%E2\%82\%80-x\%E2\%82\%80}{used linearly} (assigned grade $1$) even though it is used twice.

The \nrName{} function described here matches the ``natrec-star'' operator for linear types that we discuss in our original article \citep{icfp-version}.
While there are other ways to define the operator for the linear types modality, none would be able to solve both issues.
The reason for this is that it is the \linkFormalization[Graded.Modality.Instances.Zero-one-many]{\%E2\%8A\%9B-greatest}{largest legal definition} and we would need to assign a larger grade to solve the first problem.
We have therefore replaced the natrec-star operator with the more general \nrName{} functions.
This approach makes it possible to avoid these kinds of issues by choosing an appropriate function --- at least for instances where such a function can be defined.
As noted above it is still possible to define ``bad'' functions so there should perhaps be further restrictions on \nrName{} functions to ensure correct grade assignment.
However, we do not explore this matter here: our main goal here is correctness of erasure, and our case study shows that the constraints we impose on \nrName{} functions are sufficient to ensure this.

Elsewhere \citep{types25Natrec} we have considered an alternative method where the usage context assigned to $\natrecName$ is not computed but instead defined in terms of certain lower bounds (which are related to the characteristic inequalities we require for \nrName{} functions), but this method is \linkFormalization[Graded.Usage.Restrictions.Natrec]{Natrec-mode-supports-usage-inference}{not in general compatible} with our usage inference procedure.
There have also been various different approaches to handling inductive data types (like natural numbers) in the context of QTT \citep{qttPolyFunctor,quantitativeInductive,polyTime}.
Unlike our approach, these have the benefit of not putting the responsibility of assigning correct usage contexts on the programmer (or whoever defines the \nrName{} function).
Their drawback is that they are less flexible, either by limiting to specific modality structures or restricting when the eliminator can be applied.
For instance, \citeauthor{qttPolyFunctor} only consider eliminators which are linear in the scrutinee and \citeauthor{polyTime} require recursive calls to be erased and disallows variable uses in the zero and successor branches (in our setting these correspond to setting $\cc{r}=0$ and $\gz=\gs=\zeroConM$).
Another possible approach could be to make the assigned grade context depend on the natural number argument \citep{types25Linear}.
This feature is available to some extent in the Granule language \citep{DBLP:journals/pacmpl/OrchardLE19}, allowing for more accurate usage counting in certain situations.

As noted above, when defining an \nrName{} function one should be careful to ensure that it satisfies the intended interpretation of the modality.
Although correctness proofs beyond erasure are out of scope for this work, we will now define \nrName{} functions, that we believe to be suitable, for the example modality structures that we did not already do so for in \cref{sec:modalities}.

\paragraph*{Affine Types.}\
Recall from the usage rule for $\natrecName$ (\Cref{fig:usage}) that in $\nr{\ca p}{\cc r}{\qz}{\qs}{\qn}$, the grades $p$ and $r$ correspond to the uses of the predecessor and recursive call in the successor case while $\qz$, $\qs$, and $\qn$ correspond to the uses of the zero branch, the successor branch, and the natural number, respectively.

If $\cc r=0$, then the recursive call is unused.
Essentially, this corresponds to $\natrecName$ being used only to match on the natural number argument, meaning that either the zero branch $z$ or the successor branch $s$ will be run.
We can thus consider the usage counts of these separately and combine them using the meet operator.
In the zero case, the natural number and the zero branch are both used once, giving a combined use of $\qn + \qz$.
In the successor case, the successor branch is used once but the number of uses of the natural number depends on $\ca p$.
When $\ca p=0$, it is unused by the successor branch but used once through matching, giving it an affine use.
When $\ca p=1$ the natural number is used once through matching, providing the predecessor to be used once and when $\ca p=\omega$ it is used an unrestricted number of times.
Summarizing, the uses of the successor branch can be written $(1 \meet \ca p)\qn + \qs$ and combined with the zero branch we have $\nr{\ca p}{\cc 0}{\qz}{\qs}{\qn} = ((1 \meet \ca p)\qn + \qs) \meet (\qn + \qz)$

If $\cc r=1$, then the recursive call is used once in the step term $s$.
The zero branch will then be used once while the successor branch can be used any number of times.
The number of uses of the natural number again depends on $\ca p$.
When $\ca p=0$, it is unused in the successor case but used once through matching, providing a predecessor which is used once through matching in the recursive call.
In total, this gives one use.
When $\ca p=1$, the uses from matching are added to the use of the predecessor by the successor branch, giving an unrestricted number of uses in total.
Finally, when $\ca p=\omega$, we again have an unrestricted number of uses for the same reason.
These cases can be summarized as $\nr{\ca p}{\cc 1}{\qz}{\qs}{\qn} = (1+\ca p)\qn + \omega \qs + \qz$.

If $\cc r=\omega$, then the recursive call is used any number of times in the step term $s$.
In this case, the zero branch, successor branch and the natural number can all be used an unlimited number of times, giving a usage count of $\nr{\ca p}{\cc\omega}{\qz}{\qs}{\qn} = \omega(\qn+\qs+\qz)$.
Based on this analysis, we thus propose the following \linkFormalization[Graded.Modality.Instances.Zero-one-many]{nr}{\nrName{} function}, defined by cases on $\cc r$:
\begin{align*}
  \nr{\ca p}{\cc 0}{\qz}{\qs}{\qn} &\coloneq ((1 \meet \ca p)\qn + \qs) \meet (\qn + \qz) \\
  \nr{\ca p}{\cc 1}{\qz}{\qs}{\qn} &\coloneq (1+\ca p)\qn + \omega \qs + \qz \\
  \nr{\ca p}{\cc\omega}{\qz}{\qs}{\qn} &\coloneq \omega(\qn + \qs + \qz)
\end{align*}

\paragraph*{Linear Types.}\
The argument for affine types mostly translates also to linear types.
The only difference is the case $\cc r=0$, $\ca p=0$ where the previously affine use of the natural number in the successor case becomes unrestricted since properly affine uses cannot be expressed anymore.
However, since we defined this case of the \nrName{} function using the meet operator we still get the same \linkFormalization[Graded.Modality.Instances.Zero-one-many]{nr}{\nrName{} function} as for affine types.

\paragraph*{Linear \& Affine Types.}\
This instance combines the affine and linear types instances.
We use a similar analysis to arrive at the following \linkFormalization[Graded.Modality.Instances.Linear-or-affine]{nr}{\nrName{} function}:
\begin{align*}
  \nr{\ca p}{\cc 0}{\qz}{\qs}{\qn} &\coloneq ((1 \meet \ca p)\qn + \qs) \meet (\qn + \qz) \\
  \nr{\ca p}{\cc 1}{\qz}{\qs}{\qn} &\coloneq (1+\ca p)\qn + \omega \qs + \qz \\
  \nr{\ca p}{\cc\leone}{\qz}{\qs}{\qn} &\coloneq (\leone+\ca p)\qn + \omega \qs + \leone \qz \\
  \nr{\ca p}{\cc\omega}{\qz}{\qs}{\qn} &\coloneq \omega(\qn + \qs + \qz)
\end{align*}
Note that for $\cc r=\leone$ the zero branch is used at most once, and thus $\qz$ is multiplied by $\leone$.

\subsection{Erased matches}\label{sec:erasedMatching}
In our case study on erasure (\Cref{sec:erasure}) we identified two possible sources of erased matches and imposed restrictions to prevent problematic uses of these.
\citet{letouzey:types02} discusses similar issues in the context of extraction for Rocq, as do \citet{sozeau:coqErasure}.

The first source of erased matches is $\emptyrec{0}{A}{t}$, which allows a term $t$ to be considered unused or erasable in impossible branches.
Requiring consistent contexts allows such programs to be written but ensures that any impossible branch will never be executed when the program is run.

Few other works appear to have investigated (elimination of) the empty type in a graded setting.
\citet{DBLP:journals/corr/abs-2005-02247} do this for a system with simple types but do not allow this style of treating impossible branches as erased.
The usage rule corresponding to their system adds an arbitrary context to that of $t$ instead of scaling with some $p$ as we do.
After our original paper \citep{icfp-version} \citeauthor{liu:indistinguishability} presented a type system with something like an erased match for the empty type \citeyearpar{liu:indistinguishability}.
This kind of matching is also supported by \citet{DBLP:journals/pacmpl/Tejiscak20}.

The other source of erased matches is $\prodrecName$, which was restricted to disallow matching on erasable pairs in non-erased settings.
Previous work has studied systems both where matching on erased pairs \citep{DBLP:conf/esop/MoonEO21} or forced patterns
\citep{DBLP:journals/pacmpl/Tejiscak20} is allowed and systems where it is not \citep{choudhuryEadesEisenbergWeirich:popl21}.

Our formalization includes the ability to restrict what values the $r$ annotation on $\prodrecName$ is allowed to take.
For erasure, we can thus instantiate our system in two ways, either by allowing $r = 0$ or not, giving us the ability to explore the consequences of the choice.

As an example of the consequences of allowing or disallowing erased matches for $\prodrecName$, consider the following, weaker version of our soundness theorem, \Cref{thm:soundness}, where only the source language is considered.

\begin{linkedTheorem}[Canonicity]{Graded.Erasure.Consequences.Soundness}{Soundness\%E2\%80\%B2.Soundness.soundness-\%E2\%84\%95-only-source}\label{thm:sourceCanonicity}
    If\/ $\Gamma$ is consistent, $\wfTerm{\Gamma}{t}{\Nat}$ and $\usage{\zeroConM}{t}$ hold, and $\Prodrec{0}$ is false, then there is a natural number $n$ such that $\redSucTerm[\Gamma]{t}{\numeral{n}}$.
\end{linkedTheorem}

If the requirement about $\Prodrec{0}$ is dropped, then the theorem does not hold.
A \linkFormalization[Graded.Erasure.Consequences.Soundness]{soundness-\%E2\%84\%95-only-source-counterexample\%E2\%82\%81}{counterexample} is our example from \Cref{sec:usage}, $t \coloneq \prodrec{0}{0}{\Nat}{\var{0}}{\zero}$, which does not reduce to a numeral, but for which we have $\wfTerm{\snocCon{\epsilon}{\SigmaTypeW{0}{\Nat}{\Nat}}}{t}{\Nat}$ and $\usage{\zeroConM}{t}$, and the context $\snocCon{\epsilon}{\SigmaTypeW{0}{\Nat}{\Nat}}$ is provably consistent.
While this shows that canonicity for natural numbers is lost when erased matches are allowed, it is notable that this is \linkFormalization[Graded.Erasure.Consequences.Soundness]{soundness-\%E2\%84\%95-only-target-not-counterexample\%E2\%82\%81}{not a counterexample} to canonicity for programs compiled to the target language.
In this example the program is extracted directly to $\zero$.

Our choice to disallow erased matches for weak $\Sigma$-types in the case study was driven by provability of the fundamental lemma.
The general proof strategy consists of letting $t$ and $\erase{t}$ reduce as much as possible and use \Cref{thm:LRred}, but this strategy fails since $\prodrec{0}{q}{A}{t}{u}$ might not reduce even though $\erase{(\prodrec{0}{q}{A}{t}{u})}$ does.
In fact, the fundamental lemma \linkFormalization[Graded.Erasure.LogicalRelation.Fundamental.Counterexample]{negation-of-fundamental-lemma-with-erased-matches\%E2\%82\%81}{cannot hold if erased matches are allowed for weak $\Sigma$-types} since there are no neutral elements of type $\Nat$ in the logical relation.

Although the fundamental lemma does not hold, we conjecture that some form of canonicity for natural numbers does hold for the target language even in the presence of erased matches for weak $\Sigma$-types.
Note that the statement obtained by simply removing ``$\redsSucTerm[\Delta]{t}{\numeral{n}}$'' from the statement of \Cref{thm:soundness} might not be strong enough: the statement should arguably not be just that $\erase{t}$ reduces to a numeral, but that it reduces to the ``correct'' numeral.

\subsection{Unit types}\label{sec:unit}
The difference between the two kinds of pairs might be familiar from linear logic, where the weak and strong pairs correspond to multiplicative and additive conjunction, respectively.
Consequently, one might expect two kinds of unit types, corresponding to the multiplicative and additive unit of linear logic.
We have thus extended our formalization with such types\footnote{The strong unit type, with $\eta$-equality, was added to the formalization by Wojciech Nawrocki in 2021. We have added the weak unit type, with matching, and usage rules for both.} $\UnitX$ and unit elements $\unitX$.
As with $\Sigma$-types, $k$ can be either $\weakLabel$ or $\strongLabel$ for the weak and strong kinds respectively.
The weak unit type comes with an eliminator $\unitrecName$ in the same style as $\prodrecName$.
The strong unit type does not have an eliminator, but instead enjoys $\eta$-equality, similarly to the strong $\Sigma$-type.
The typing and reduction rules for the unit types can be found in \Cref{fig:unitTyping}.

\begin{figure}
  \begin{mathpar}
    \inferrule{\wfCon{\Gamma}}{\wfTerm{\Gamma}{\UnitX[\cc k]}{\U}} \and
    \inferrule{\wfCon{\Gamma}}{\wfTerm{\Gamma}{\unitX[\cc k]}{\UnitX[\cc k]}} \and
    \inferrule{\wfType{\snocCon{\Gamma}{\UnitX[\cc \weakLabel]}}{A}\\\wfTerm{\Gamma}{t}{\UnitX[\cc \weakLabel]}\\\wfTerm{\Gamma}{u}{\subst{A}{\unitX[\cc \weakLabel]}}}{\wfTerm{\Gamma}{\unitrec{p}{q}{A}{t}{u}}{\subst{A}{t}}} \and
    \inferrule{\wfTerm{\Gamma}{t}{\UnitX[\cc \strongLabel]}\\\wfTerm{\Gamma}{u}{\UnitX[\cc \strongLabel]}}{\eqTerm{\Gamma}{t}{u}{\UnitX[\cc \strongLabel]}}\and
    \inferrule{\eqType{\snocCon{\Gamma}{\UnitX[\cc \weakLabel]}}{A}{A'}\\\eqTerm{\Gamma}{t}{t'}{\UnitX[\cc \weakLabel]}\\\eqTerm{\Gamma}{u}{u'}{\subst{A}{\unitX[\cc \weakLabel]}}}{\eqTerm{\Gamma}{\unitrec{\ca p}{\cb q}{A}{t}{u}}{\unitrec{\ca p}{\cb q}{A'}{t'}{u'}}{\subst{A}{t}}} \and
    \inferrule{\wfType{\snocCon{\Gamma}{\UnitX[\cc \weakLabel]}}{A}\\\wfTerm{\Gamma}{u}{\subst{A}{\unitX[\cc \weakLabel]}}}{\eqTerm{\Gamma}{\unitrec{p}{q}{A}{\unitX[\cc \weakLabel]}{u}}{u}{\subst{A}{\unitX[\cc \weakLabel]}}} \and
    \inferrule{\wfType{\snocCon{\Gamma}{\UnitX[\cc \weakLabel]}}{A}\\\wfTerm{\Gamma}{u}{\subst{A}{\unitX[\cc \weakLabel]}}\\\redTerm{\Gamma}{t}{t'}{\UnitX[\cc \weakLabel]}}{\redTerm{\Gamma}{\unitrec{\ca p}{\cb q}{A}{t}{u}}{\unitrec{\ca p}{\cb q}{A}{t'}{u}}{\subst{A}{t}}} \and
    \inferrule{\wfType{\snocCon{\Gamma}{\UnitX[\cc \weakLabel]}}{A}\\\wfTerm{\Gamma}{u}{\subst{A}{\unitX[\cc \weakLabel]}}}{\redTerm{\Gamma}{\unitrec{\ca p}{\cb q}{A}{\unitX[\cc \weakLabel]}{u}}{u}{\subst{A}{\unitX[\cc \weakLabel]}}}
  \end{mathpar}
  \Description{Extensions of the judgements for well-formed types, $\wfType{\Gamma}{A}$, terms, $\wfTerm{\Gamma}{t}{A}$, and term reduction, $\redTerm{\Gamma}{t}{u}{A}$, with the inclusion of unit types.}
  \caption{Extensions of the judgements for well-formed types, \linkFormalization[Definition.Typed]{_\%E2\%8A\%A2_}{$\wfType{\Gamma}{A}$}, terms, \linkFormalization[Definition.Typed]{_\%E2\%8A\%A2_\%E2\%88\%B7_}{$\wfTerm{\Gamma}{t}{A}$}, and term reduction, \linkFormalization[Definition.Typed]{_\%E2\%8A\%A2_\%E2\%87\%92_\%E2\%88\%B7_}{$\redTerm{\Gamma}{t}{u}{A}$}, with the inclusion of unit types.}\label{fig:unitTyping}
\end{figure}

More interesting is perhaps what usage rules should be assigned to these.
For the type itself we assign the zero usage context $\zeroConM$ as we did for the other base types.
For the unit elements we might expect that the rules differ between the two kinds in the same way as in linear logic or as the $\Sigma$-types do.
We take the approach that the unit elements should serve as units for their respective pairs in terms of usage.
That is, terms like $\pairX{t}{\unitX}$ and $t$ should have the same resource usage.

For the weak unit type, this means that we assign the context $\zeroConM$ since it is the identity for addition.
The rule for $\unitrecName$ follows the pattern of $\prodrecName$, allowing the uses of the scrutinee to be scaled by a given grade annotation corresponding to its number of uses, see \Cref{fig:unitUsage}.
In particular, it allows an erased match where no uses of the unit element are counted.
As for pairs, we allow preventing this kind of match (as well as matches with other grade annotations) through a configurable side condition.

For the strong unit element, our approach might suggest that we should assign a context that is the identity for meet.
This would mean that we should assign the greatest usage context, but our modalities do not necessarily include a greatest element (the linear types instance is an example).
Instead, we note that due to the idempotence of meet, we could achieve our goal if the usage context assigned to $\unitS$ above is the same as the one assigned to $t$.
We thus allow the strong unit element to be used in any context.
In a sense, $\unitS$ acts as a ``sink'', allowing a program to ``consume'' any variable an arbitrary number of times.
These usage rules are summarized in \Cref{fig:unitUsage}.
Like for pairs the rules for the unit elements match those of linear logic \citep{linearLogic}, with the weak unit corresponding to the multiplicative unit ($1$) which can only be derived from no assumptions, while the strong unit corresponds to the additive unit ($\top$) which is derivable under any assumptions.

\begin{figure}
\begin{mathpar}
  \inferrule{ }{\usage{\zeroConM}{\UnitX[\cc k]}} \and
  \inferrule{ }{\usage{\zeroConM}{\unitX[\cc \weakLabel]}} \and
  \inferrule{\usage{\cd\gamma}{t}\\\usage{\cf\delta}{u}\\\usage{\snocCon{\eta}{\cb q}}{A}}{\usage{\ca{p}\cd\gamma + \cf\delta}{\unitrec{\ca p}{\cb q}{A}{t}{u}}} \; \Unitrec{\ca p} \and
  \inferrule{ }{\usage{\gamma}{\unitX[\cc \strongLabel]}}
\end{mathpar}
\Description{Extensions of the usage relation, $\usage{\gamma}{t}$, with the inclusion of unit types.}
\caption{Extensions of the usage relation, \linkFormalization[Graded.Usage]{_\%E2\%96\%B8\%5B_\%5D_}{$\usage{\gamma}{t}$}, with the inclusion of unit types.}\label{fig:unitUsage}
\end{figure}

The weak unit type allows us to \linkFormalization[Definition.Untyped.Empty]{emptyrec-sink}{encode a variant of $\emptyrecName$} which is \linkFormalization[Graded.Derived.Empty]{\%E2\%96\%B8emptyrec-sink}{well-resourced under any context} (if its arguments are well-resourced).
This variant is more flexible than the original one since it allows variables to be assigned arbitrary uses as opposed to being limited by the uses in the scrutinee.
For instance, with this variant variables may be considered to be linear despite being erased in the scrutinee.
This is perhaps more natural than our original usage rule since it should be fine to assign any number of uses to code that will never run.

A consequence of our usage rule for the strong unit element is that the inference (and decision) procedure for usage we described in \Cref{sec:usage} no longer works as it is based on the syntax containing sufficient grade annotations.%
\footnote{If $0$ is a largest grade, then one can infer the grade vector $\zeroConM$ for $\unitS$.}
To get the usage rule $\usage{\gamma}{\unitS}$ and still support usage inference, one could perhaps annotate $\unitS$ with the vector of resources to be consumed.
We did not pursue this approach further, as it would seemingly require a large change to our formalization.
In particular, what should happen with the annotations when one applies a substitution?
Perhaps one could pass around suitable substitution matrices.

Apart from usage inference, the results we have presented above hold also in the presence of these unit types.
The logical relation for types is extended in the expected way:
\begin{equation*}
  \inferrule*[left={$\rUnitX{\cc k}$},vcenter]{\redsType{\Gamma}{A}{\UnitX[\cc k]}}{\typeCode{\Gamma}{A}}
\end{equation*}
Similarly, the target language is extended with a unit element and eliminator and corresponding reduction rules (see the formalization for details).
The extraction function is updated as follows:
\begin{mathpar}
  \erase{\UnitX} \coloneqq \eraseType\and
  \erase{\unitX} \coloneq \unit\and
  \erase{\unitrec{\ca 0}{q}{A}{t}{u}} \coloneq \erase{u}\and
  \erase{\unitrec{\ca\omega}{q}{A}{t}{u}} \coloneq \unitrecT{\erase{t}}{\erase{u}}
\end{mathpar}
Erased matches are thus removed, just like for weak $\Sigma$-types.

The erasure logical relation is extended such that $\eTerm{t}{v}{A}{\rUnitX{\cc k}}$ holds iff $\redsTerm{\Delta_0}{t}{\unitX[\cc k]}{\UnitX[\cc k]}$ and $\redsT{v}{\unit}$, ensuring that both terms reduce to unit elements.
The fundamental lemma and soundness of erasure can be proven as before with the caveat that erased matches for $\unitrecName$ should be disallowed where it was disallowed for $\prodrecName$.
The discussion about erased matches for $\prodrecName$ in \Cref{sec:erasedMatching} applies also to $\unitrecName$.
In particular, one can construct a similar \linkFormalization[Graded.Erasure.Consequences.Soundness]{soundness-\%E2\%84\%95-only-source-counterexample\%E2\%82\%85}{counterexample} to \Cref{thm:sourceCanonicity} when erased matches for $\unitrecName$ are allowed.

\subsection{Information flow interpretation}\label{sec:security}
The erasure lattice $\omega \leq 0$ also serves as a two-point
information flow lattice $\L \leq \H$ where publicly
available resources are labelled with $\L$ and private (confidential)
resources are labelled with $\H$.
A program $t$ is viewed from the public perspective, and
a context $\gamma$ with $\usage \gamma t$ describes how it accesses its resources
(its free variables).
The law $\L + \H = \L$ expresses that a resource that is accessed both
publicly and privately must be publicly available.
Our fundamental lemma for the erasure logical relation (\Cref{thm:fundamental}) implies a form of \linkFormalization[Graded.Erasure.Consequences.Non-interference]{non-interference}{non-interference}:
the result of computing a natural number does not depend on the values of $\H$-labelled resources.

In a two-sided formulation $\gamma\Gamma \vdash t : pA$ one can view
programs from perspectives $p$ other than $\L = 1$.
Depending on what rules are used one may have that a variable with index $x$ can be accessed if and only if $\gamma(x) \leq p$.
For $p = \H$ this means that any variable can be accessed.

One can perhaps
simulate the two-sided approach by shifting the perspective
from $1$ to $p$ by \emph{dividing} $\gamma$ by $p$.
This happens pointwise, so $(\gamma/p)(x) = \gamma(x)/p$.
Let us say that a modality structure \emph{supports division by $q$} if there is a function $-/q$ such that the Galois connection $\forall p, r. \ p/q \leq r \iff p \leq q \cdot r$ holds.
Note that $p/q$ is the \linkFormalization[Graded.Modality.Properties.Division]{\%2F-least-\%E2\%89\%A4\%C2\%B7}{least} $r$ such that $p \leq q \cdot r$, and that $-/q$ is \linkFormalization[Graded.Modality.Properties.Division]{\%2F-monotone\%CB\%A1\%E2\%80\%B2}{monotone}.
The erasure instance supports division by $\L$, with $p/\L = p$, and by $\H$, with $p/\H = \L$.
The latter equality entails that all resources are accessible from a private perspective.

This shift of perspective can be carried out in finer information flow lattices.
Recall that bounded, distributive lattices with decidable equality can be seen as modality structures where $1$ is the least element, $0$ is the greatest element, addition is meet, and multiplication is join.
One example of such a lattice is the total order \linkFormalization[Graded.Modality.Instances.Information-flow]{L\%E2\%89\%A4M\%E2\%89\%A4H}{$1 = \L \leq \M \leq \H = 0$},
where we have added a \emph{medium} clearance to the low and high ones.
It is straightforward to derive the following division laws in such lattices:
\linkFormalization[Graded.Modality.Properties.Division]{\%2F\%F0\%9D\%9F\%99\%E2\%89\%A1}{$p/1 = p$} and
\linkFormalization[Graded.Modality.Properties.Division]{\%2F\%F0\%9D\%9F\%98\%E2\%89\%A1\%F0\%9D\%9F\%99}{$p/0 = 1$} and
\linkFormalization[Graded.Modality.Properties.Division]{\%2F\%E2\%89\%A1\%F0\%9D\%9F\%99}{$p/p = 1$} and
\linkFormalization[Graded.Modality.Properties.Division]{\%F0\%9D\%9F\%99\%2F\%E2\%89\%A1\%F0\%9D\%9F\%99}{$1/p = 1$} and
\linkFormalization[Graded.Modality.Properties.Division]{\%F0\%9D\%9F\%98\%2F\%E2\%89\%A1\%F0\%9D\%9F\%98}{$0/p = 0$}.
 The last three are given under the assumption that division by $p$ is supported, and for the last one we additionally assume that $p \not= 0$ and that the zero-product property holds.
For the lattice $\L \leq \M \leq \H$ we get that $\L/\M = \L$ and $\M/\M = \L$ and $\H/\M = \H$, so shifting the perspective to $\M$ makes $\M$-labelled resources available.

\section{Extension: modes and graded \texorpdfstring{$\Sigma$}{Σ}-types}\label{sec:modes}
Let us now present an extension of the theory with support for
\emph{graded Σ-types}:
The type $\GradedSigmaTypeX{p}{q}{A}{B}$ is a variant of
$\SigmaTypeX{q}{A}{B}$ where the first component, of type $A$, has
grade $p$.
If the erasure instance is used then $\GradedSigmaTypeX{0}{q}{A}{B}$
is a $\Sigma$-type with an erased first component.
This kind of type can be used to have an erased component, such as a
proof of an invariant, in a data structure.

Note that it might not make sense to apply $\fstName$ to a value of
type, say, $\GradedSigmaTypeS{0}{0}{\Nat}{\Nat}$: if the first
component is erased then there is nothing to return.
So, in order to support graded $\Sigma$-types we follow
\citet{DBLP:conf/lics/Atkey18} and switch to a usage relation with one
of two \linkFormalization[Graded.Mode.Instances.Zero-one]{Mode}{\emph{modes}}: $\modeZero$ (erased/compile-time) or $\modeOne$
(present/run-time).
The idea is that $\fstName$ is allowed for the graded $\Sigma$-type
$\GradedSigmaTypeS{0}{q}{A}{B}$ only if the mode is $\modeZero$,
whereas there are no extra restrictions for
$\GradedSigmaTypeS{p}{q}{A}{B}$ when $p \neq 0$.

In this section we work with modality structures with well-behaved zeros
(\Cref{def:well-behaved-zero}), \ie, we have $0 \neq 1$,
addition, meet, and $\nrName$ are positive, and the zero-product property holds.
We require that $0 \neq 1$ to ensure that the modes $\modeZero$ and
$\modeOne$ correspond to different grades.

As mentioned above, the instances for erasure, affine types, linear
types, and linear or affine types discussed in \Cref{sec:modalities}
all have well-behaved zeros, and we have proved a subject reduction
lemma for arbitrary such instances
(\Cref{thm:subject-reduction-modes}).
However, note that our main soundness theorem
(\Cref{thm:soundness-modes}) is focused on erasure.
Thus, we cannot claim that the definitions below make sense in all contexts
for all instances with well-behaved zeros.

Modes can be \linkFormalization[Graded.Mode.Instances.Zero-one]{\%E2\%8C\%9C_\%E2\%8C\%9D}{turned
  into} grades: $\modeToGrade{\modeZero} = 0$
and $\modeToGrade{\modeOne} = 1$.
Grades can also be
\linkFormalization[Graded.Mode.Instances.Zero-one]{\%E2\%8C\%9E_\%E2\%8C\%9F}{translated} to modes
($\gradeToMode{0} = \modeZero$, and $\gradeToMode{p} = \modeOne$ if
$p \neq 0$), and modes can be
\linkFormalization[Graded.Mode]{IsMode._\%E1\%B5\%90\%C2\%B7_}{scaled} by values
of type $\mathcal{M}$ ($\modeGradeMode{\modeZero}{p} = \modeZero$ and
$\modeGradeMode{\modeOne}{p} = \gradeToMode{p}$).

\linkFormalization[Definition.Untyped]{Term}{The syntax} is modified a
little to support $\GradedSigmaTypeX{p}{q}{A}{B}$: the projections
$\fstName$ and $\sndName$, the pair constructor, and $\prodrecName$
are all annotated with a grade corresponding to $p$ (in addition to
any prior annotations).
The \linkFormalization[Definition.Typed]{\%E2\%8A\%A2_}{type system} and \linkFormalization[Definition.Typed]{_\%E2\%8A\%A2_\%E2\%87\%92_\%E2\%88\%B7_}{reduction relations} are mostly unchanged, except to
ensure that the $p$ annotations match when appropriate --- see the
formalization for details.
However, the usage relation is modified substantially.
Its definition is inspired by the type system presented by
\citet{DBLP:conf/lics/Atkey18}.

\begin{linkedDefinition}[Moded grading]{Graded.Usage}{_\%E2\%96\%B8\%5B_\%5D_}
  Given a usage context $\gamma$, a mode $m$ and a term $t$, we say
  that $t$ is \emph{well-resourced} with respect to $\gamma$ and $m$
  if $\usageM{\gamma}{m}{t}$, where the relation $\usageMName$ is
  defined inductively in \Cref{fig:moded-grading}.
  \label{def:usage-with-mode}
\end{linkedDefinition}
\begin{figure}
  \begin{mathpar}
    \inferrule{ }{\usageM{\zeroConM}{\cc m}{\U}}
    \and
    \inferrule{ }{\usageM{\zeroConM}{\cc m}{\Nat}}
    \and
\TOP{
    \inferrule{ }{\usageM{\zeroConM}{\cc m}{\Unit}}
    \and
}
    \inferrule{ }{\usageM{\zeroConM}{\cc m}{\Empty}}
    \and
    \inferrule{
        \usageM{\gamma}{\modeGradeMode{\cc m}{\ca p}}{A} \\
        \usageM{\snocCon{\delta}{\modeToGrade{\cc m}q}}{\cc m}{B}
    }{\usageM{p\gamma + \delta}{\cc m}{\PiType{p}{q}{A}{B}}}
    \and
    \inferrule{
        \usageM{\gamma}{\modeGradeMode{\cc m}{\ca p}}{A} \\
        \usageM{\snocCon{\delta}{\modeToGrade{\cc m}q}}{\cc m}{B}
    }{\usageM{\ca p\gamma + \delta}{\cc m}{\GradedSigmaTypeX{\ca p}{q}{A}{B}}}
    \and
    \inferrule{ }{\usageM{\modeToGrade{\cc m}\varConM{i}}{\cc m}{\var{i}}}
    \and
    \inferrule{\usageM{\snocCon{\gamma}{\modeToGrade{\cc m}p}}{\cc m}{t}}{%
      \usageM{\gamma}{\cc m}{\lam{p}{t}}}
    \and
    \inferrule{
            \usageM{\gamma}{\cc m}{t} \\
            \usageM{\delta}{\modeGradeMode{\cc m}{\ca p}}{u}
    }{\usageM{\gamma + p \delta}{\cc m}{\app{t}{p}{u}}}
    \and
    \inferrule{
        \usageM{\gamma}{\modeGradeMode{\cc m}{\ca p}}{t} \\
        \usageM{\delta}{\cc m}{u}
    }{\usageM{\ca p\gamma + \delta}{\cc m}{\pairMW{t}{u}{\ca p}}}
    \and
    \inferrule{
        \usageM{\gamma}{\modeGradeMode{\cc m}{\ca p}}{t} \\
        \usageM{\delta}{\cc m}{u}
    }{\usageM{\ca p\gamma\meet\delta}{\cc m}{\pairMS{t}{u}{\ca p}}}
    \and
    \inferrule{\usageM{\gamma}{\cc m}{t}}{%
      \usageM{\gamma}{\cc m}{\fstM{\ca p}{t}}}
    \; \modeGradeMode{\cc m}{\ca p} \leq \modeToGrade{\cc m}
    \and
    \inferrule{\usageM{\gamma}{\cc m}{t}}{%
      \usageM{\gamma}{\cc m}{\sndM{\ca p}{t}}}
    \and
    \inferrule{
        \usageM{\gamma}{\modeGradeMode{\cc m}{r}}{t} \\
        \usageM{\snocCon{\snocCon{\delta}{\modeToGrade{\cc m}r\ca p}}{\modeToGrade{\cc m}r}}{\cc m}{u} \\
        \usageM{\snocCon{\eta}{0}}{\modeZero}{A}
    }
    {\usageM{r \gamma + \delta}{\cc m}{\prodrecM{r}{\ca p}{q}{A}{t}{u}}}
    \; \Prodrec{r}
    \and
    \inferrule{ }{\usageM{\zeroConM}{\cc m}{\zero}}
    \and
    \inferrule{\usageM{\gamma}{\cc m}{t}}{%
      \usageM{\gamma}{\cc m}{\suc{t}}}
    \and
    \inferrule{
        \usageM{\gamma}{\cc m}{z} \\
        \usageM{\snocCon{\snocCon{\delta}{\modeToGrade{\cc m}p}}
                  {\modeToGrade{\cc m}r}}
          {\cc m}{s} \\
        \usageM{\eta}{\cc m}{n} \\
        \usageM{\snocCon{\theta}{0}}{\modeZero}{A}
    }
    {\usageM{\nr{p}{r}{\gamma}{\delta}{\eta}}
       {\cc m}{\natrec{p}{q}{r}{A}{z}{s}{n}}
    }
    \and
    \inferrule{
        \usageM{\gamma}{\modeGradeMode{\cc m}{\ca p}}{t} \\
        \usageM{\delta}{\modeZero}{A}
    }
    {\usageM{p\gamma}{\cc m}{\emptyrec{p}{A}{t}}}
    \; \Emptyrec{p}
    \and
\TOP{
    \inferrule{ }{\usageM{\zeroConM}{\cc m}{\unit}}
    \and
}
    \inferrule{\usageM{\gamma}{\cc m}{t}}{%
      \usageM{\delta}{\cc m}{t}} \; \delta\le\gamma
  \end{mathpar}
  \Description{Moded grading $\usageM{\gamma}{m}{t}$. New variable
    occurrences are highlighted in colour.}
  \caption{Moded grading \linkFormalization[Graded.Usage]{_\%E2\%96\%B8\%5B_\%5D_}{$\usageM{\gamma}{m}{t}$}. New variable
    occurrences are highlighted in colour.}\label{fig:moded-grading}
\end{figure}

The variable rule uses the usage context $\modeToGrade{m}\varConM{i}$.
If $m$ is $\modeOne$, then we get the same context as before, but if
$m$ is $\modeZero$, then we get
$\modeToGrade{\modeZero}\varConM{i} = \zeroConM$, so the variable is
not counted.
In the antecedent of the rule for $\lambda$-abstractions the grade of
variable $0$ is $\modeToGrade{m}p$: if $m$ is $\modeZero$, then
$\modeToGrade{m}p = 0$, which matches the variable rule's grade
context for $\modeZero$.
Multiplication by $\modeToGrade{m}$ is used in this way also in other
rules.

Note that $\modeGradeMode{m}{p}$ is $\modeZero$ exactly when
$m = \modeZero$ or $p = 0$.
This construction is used when we want to ensure that things are
checked in the erased mode $\modeZero$ if either the mode is already
$\modeZero$, or some argument is erased, for instance for
$\PiType{0}{q}{A}{B}$, $\GradedSigmaTypeX{0}{q}{A}{B}$ and
$\app{t}{0}{u}$.

The usage rules given above are partly based on Agda's rules for
erasure \citep{abel-danielsson-vezzosi-erased-univalence}.
Agda supports cubical type theory, which imposes restrictions on what
rules one can use for erasure (see \Cref{sec:also-in-types}).
We have tried to be compatible with the rules of Agda.
In particular, the rules for $\Pi$ and $\Sigma$ are based on those of
Agda.
In Agda $p$ and $q$ are required to be equal, and the formalization
has a parameter which makes it possible to enforce this requirement.
When $p$ and $q$ are both $0$ we get the following derived rule for $\Sigma$ (and something similar for $\Pi$):
\begin{mathpar}
  \inferrule{
      \usageM{\gamma}{\modeZero}{A} \\
      \usageM{\snocCon{\delta}{0}}{\cc m}{B}
  }{\usageM{\delta}{\cc m}{\GradedSigmaTypeX{0}{0}{A}{B}}}
\end{mathpar}
Note that $A$ is checked in the erased mode $\modeZero$.
The rule is motivated by Cubical Agda's \emph{transport} primitive, which is type-driven: for a $\Sigma$-type with an erased first component transport applied to a pair is equal to a pair containing two applications of transport, and the domain $A$ is used in erased positions in both.
We have a similar situation for $\Pi$-types.

One could imagine making $\Pi$- and $\Sigma$-types even more general, with four grades: one for terms (lambdas, pairs and so on) and three for the different uses of ``$p$'' and ``$q$'' in the usage rules for $\Pi$ and $\Sigma$.
However, we have refrained from this.

The term $\fstM{p}{t}$ can only be used if the side condition $\modeGradeMode{m}{p} \le \modeToGrade{m}$ is satisfied so if $p = 0$, then the mode must be
$\modeZero$.
This condition is used in the proof of
\Cref{thm:subject-reduction-modes}.
Note that for the erasure, affine types, linear types, and linear or
affine types modalities, the only grade that is not bounded by $1$ is
$0$.

The first component of a pair of type $\GradedSigmaTypeX{p}{q}{A}{B}$
is associated with a grade $p$, and correspondingly the rules for the
pair constructors include scaling by $p$ in their conclusions.
Scaling by $p$ is also used in the rule for $\prodrecName$: note that
if $p$ is $0$, then there must be no (counted) uses of the variable
corresponding to the first component in $u$.

The rules for $\prodrecName$, $\natrecName$ and $\emptyrecName$ still
include assumptions about the motives (``$A$''), but now these
assumptions use the mode $\modeZero$, and in the case of
$\prodrecName$ and $\natrecName$ the last grade in the usage context
is $0$: the grade $q$ is no longer used.
Note that $\usageM{\zeroConM}{\modeZero}{A}$ holds
\linkFormalization[Graded.Usage.Restrictions.Satisfied]{\%F0\%9D\%9F\%98\%E1\%B6\%9C\%E2\%96\%B8\%5B\%F0\%9D\%9F\%98\%E1\%B5\%90\%5D\%E2\%87\%94}{exactly}
when $\Prodrec{r}$ and $\Emptyrec{r}$ holds for all subterms in $A$ of the form
$\prodrecM{r}{p}{q}{B}{t}{u}$ and $\emptyrec{r}{B}{t}$ respectively.

We can prove a substitution lemma for the theory with graded $\Sigma$-types,
and using that we obtain the following variant of subject reduction
for usage:
\begin{linkedTheorem}[Moded subject reduction]{Graded.Reduction}{Subject-reduction.usagePresTerm}
  \label{thm:subject-reduction-modes}
  If $\usageM{\gamma}{m}{t}$ and $\redTerm{\Gamma}{t}{u}{A}$ then
  $\usageM{\gamma}{m}{u}$.
\end{linkedTheorem}

The extraction function from \Cref{sec:erasure} is
\linkFormalization[Graded.Erasure.Extraction]{erase}{modified} for some term
formers corresponding to graded $\Sigma$-types.
Erased first components are erased entirely, and $\fstM{0}{t}$ is
replaced by $\loopT$.
If $r \ne 0$ then $\prodrecM{r}{0}{q}{A}{t}{u}$ is extracted to $\appT{(\subst{(\lamT{\erase{u}})}{\loopT})}{\erase{t}}$, where $\loopT$ is substituted for the (erased) first component.
The updated cases are as follows (here $\omega$ stands for any grade distinct
from $0$):
\begin{displaymath}
  \defaultcolumn{@{}l@{}}
  \begin{pmboxed}
    \>\erase{\pairM{t}{u}{\ca 0}} \>\coloneqq \erase{u}
    \>[][!@{\quad\;\;}l@{}]\erase{\pairM{t}{u}{\ca\omega}} \>\coloneqq
      \nonstrict{\ensuremath{\pair{\erase{t}}{\erase{u}}}} /
      \strict{\ensuremath{\appT{\appT{(\lamT{\lamT{\pair{\var{1}}{\var{0}}}})}{\erase{t}}}{\erase{u}}}} \\
    \>\erase{(\fstM{\ca 0}{t})} \>\coloneqq\loopT
    \>\erase{(\fstM{\ca\omega}{t})} \>\coloneqq\fst{\erase{t}} \\
    \>\erase{(\sndM{\ca 0}{t})} \>\coloneqq\erase{t}
    \>\erase{(\sndM{\ca\omega}{t})} \>\coloneqq\snd{\erase{t}} \\
    \>\erase{(\prodrecM{\cc 0}{p}{q}{A}{t}{u})} \>\coloneqq \substtwo{\erase{u}}{\loopT}{\loopT}
    \>\erase{(\prodrecM{\cc\omega}{\ca 0}{q}{A}{t}{u})} \>\coloneqq \appT{(\subst{(\lamT{\erase{u}})}{\loopT})}{\erase{t}} \\
    \>\>\>\erase{(\prodrecM{\cc\omega}{\ca\omega}{q}{A}{t}{u})} \>\coloneqq \prodrecT{\erase{t}}{\erase{u}}
  \end{pmboxed}
\end{displaymath}
As before, \strict{red} and \nonstrict{green} backgrounds refer to the strict and non-strict variants of the target language, respectively.

With these definitions in place we can state the main soundness
theorem for the theory with graded $\Sigma$-types (the statement is identical
to that of \Cref{thm:soundness}).
We do not include a proof here but refer to the formalization for
details.
\begin{linkedTheorem}[Operational erasure soundness]{Graded.Erasure.Consequences.Soundness}{Soundness\%E2\%80\%B2.Soundness.soundness-\%E2\%84\%95}
  \label{thm:soundness-modes}
  Let $\Delta$ be a context.
  If\/ $\Prodrec{0}$ is false or $\Delta$ is empty,
  $\Delta$ is consistent if $\Emptyrec{0}$ is true,
  and furthermore
  $\wfTerm{\Delta}{t}{\Nat}$ and $\usageM{\zeroConM}{\modeOne}{t}$, then
  there is a natural number $n$ such that
  $\redsSucTerm[\Delta]{t}{\numeral{n}}$ and $\redsSucT{\erase{t}}{\numeral{n}}$.
\end{linkedTheorem}

Note that the reduction relation does not include $\eta$-expansion.
We have proved that there is a type- and
resource-preserving procedure that takes a well-typed, well-resourced
term to one of its $\eta$-long normal forms.
However, there are some restrictions.
The linear identity function $\lamN{1}{x}{x}$ of type
$\PiType{1}{r}{(\GradedSigmaTypeS{p}{q}{\Nat}{\Nat})}{(\GradedSigmaTypeS{p}{q}{\Nat}{\Nat})}$
is definitionally equal to the $\eta$-long normal form
$\lamN{1}{x}{\pairMS{\fstM{p}{x}}{\sndM{p}{x}}{p}}$, which is
well-resourced in the empty context
\linkFormalization[Graded.Reduction.Zero-one]{\%CE\%B7-long-nf-for-id\%E2\%87\%94\%E2\%89\%A1\%F0\%9D\%9F\%99\%E2\%8A\%8E\%E2\%89\%A1\%F0\%9D\%9F\%98}{if
  and only if} $p = 1$ or $1 \leq 0 = p$.
We have \linkFormalization[Graded.FullReduction]{fullRedTerm}{proved} that
well-typed, well-resourced terms have $\eta$-long normal forms if the
theory is restricted so that, if $\GradedSigmaTypeSWithoutArgs{p}{q}$
is allowed, then $p = 1$ or $1 \leq 0 = p$.
For the
\linkFormalization[Graded.Modality.Instances.Erasure.Properties]{Zero-one.full-reduction-assumptions}{erasure}
instance these conditions always hold, without any restrictions.
For the instance of
\linkFormalization[Graded.Modality.Instances.Affine]{Zero-one.full-reduction-assumptions}{affine
  types} the conditions hold if
$\GradedSigmaTypeSWithoutArgs{\omega}{q}$ is disallowed.
And finally the conditions hold for the instances of
\linkFormalization[Graded.Modality.Instances.Linearity]{Zero-one.full-reduction-assumptions}{linear
  types} and
\linkFormalization[Graded.Modality.Instances.Linear-or-affine]{Zero-one.full-reduction-assumptions}{linear
  or affine} types if $\GradedSigmaTypeSWithoutArgs{p}{q}$
is only allowed when $p = 1$.
Note that there are no restrictions for
$\GradedSigmaTypeWWithoutArgs{p}{q}$.

The graded $\Sigma$-types discussed here are similar to the tensor products
presented by \citet{DBLP:conf/lics/Atkey18}, but there are some
differences.
\citeauthor{DBLP:conf/lics/Atkey18} has a single tensor product/$\Sigma$-type
with a single usage annotation.
In his theory the projections $\fstName$ and $\sndName$ are allowed
only in the erased fragment (when the mode is $\modeZero$), whereas a
construction similar to $\prodrecName$ is always allowed (but there is
nothing like the ``third grade $r$'', and erased matches are not
supported).
The usual $\eta$-equality rule for $\Sigma$-types is available for tensor products
only in the erased fragment: \citeauthor{DBLP:conf/lics/Atkey18} does
not separate usage from typing, so it is straightforward to restrict
an equality rule in this way.

\section{Conclusions}\label{sec:conclusions}
We have presented a dependently typed core language with grades
for variable usage, generic over a semiring-like modality structure.  The language
features $\Pi$, weak and strong $\Sigma$, a universe, an empty type,
and natural numbers.  We have formalized
the meta-theory up to normalization and decidability of definitional
equality in Agda, including standard results such as substitution
lemmata, and consistency, building on the work of \citet{DBLP:journals/pacmpl/0001OV18}.  We further added a usage
calculus and showed its correctness via a subject reduction result.
As an instance of our generic usage calculus, we studied modality structures able to track erasure and proved their soundness with respect to program extraction by a logical relations argument.

Some loose ends that could be investigated in future work:
\begin{itemize}

\item Utilize the formalization to implement a verified type checker for graded modal type theory, perhaps along the lines of McTT~\cite{pientka:mctt}.  This type checker would be provably terminating and produce (erased) derivations of well-typedness.

\item Prove that extraction is sound also in the presence of erased
  matches for weak $\Sigma$-types.

\item Further investigation of grade assignment for natural numbers, or recursive types more generally.

\item Add grade inference, following
  \citet{DBLP:journals/pacmpl/Tejiscak20}.

\item Investigate the interaction of modalities with meta-variables
  used for type reconstruction in real-world dependently typed
  languages such as Agda and Idris.
  Implementations already exist, but theoretical work is still lacking.

\end{itemize}

\section*{Acknowledgements}
We thank the anonymous reviewers of this and the previous version of this article for their valuable comments and suggestions.
We also thank
Naïm Camille Favier,
Eve Geng,
Gaëtan Gilbert,
Ondřej Kubánek,
Wojciech Nawrocki,
Joakim Öhman, and
Andrea Vezzosi
for their contributions to the formalization.

Andreas Abel and Oskar Eriksson acknowledge support by
\emph{Vetenskapsr\aa{}det}
(the Swedish Research Council) via project 2019-04216
\emph{Modal typteori med beroende typer} (Modal Dependent Type Theory).
Oskar Eriksson additionally acknowledges support by \emph{Knut and Alice Wallenberg Foundation} via project 2019.0116.
Nils Anders Danielsson acknowledges support from Vetenskapsrådet (2023-04538).

\bibliography{references}

\end{document}